\documentclass[sigconf, nonacm]{acmart}
\usepackage{tikz}
\usepackage{comment}
\usepackage{caption}
\usepackage{subcaption}

\usepackage{siunitx}
\usepackage{braket}

\usepackage{xcolor}
\usepackage{tabularx, makecell, linegoal}

\usepackage{ifthen}

\newboolean{ShowComments}
\setboolean{ShowComments}{true}  
\ifthenelse{\boolean{ShowComments}}%
	{
		\newcommand{\ColorComment}[3]{%
				{\colorbox{#1}{\color{white}   \textsf{\textbf{#2}}} \textcolor{#1}{#3}}}

	}%
	{
		\newcommand{\ColorComment}[3]{}

	}%

\definecolor{sasakicolor}{RGB}{120, 60, 180}
\definecolor{shotacolor}{RGB}{0, 0, 255}
\definecolor{rdvcolor}{RGB}{0,128,0}\newcommand{\rdv}[1]{\ColorComment{rdvcolor}{rdv}{#1}}
\definecolor{michalcolor}{RGB}{255,127,80}
\definecolor{amincolor}{RGB}{96, 189, 
239}\newcommand{\amin}[1]{\ColorComment{amincolor}{amin}{#1}}
\definecolor{sakumacolor}{RGB}{170, 
80, 200}\newcommand{\sakuma}[1]{\ColorComment{sakumacolor}{sakuma}{#1}}
\definecolor{andrewcolor}{RGB}{255, 
80, 200}

\def\u#1{_{\rm #1}}

\usepackage{CJKutf8}

\AtBeginDocument{%
  }

\setcopyright{acmlicensed}
\copyrightyear{2025}
\acmYear{2025}
\acmDOI{XXXXXXX.XXXXXXX}
\acmConference[ASPLOS '26]{the ACM International Conference on Architectural Support for Programming Languages and Operating Systems}{March 22-26, 2026}{Pittsburgh, USA}
\acmISBN{978-1-4503-XXXX-X/2018/06}

\begin{document}

\title[]{Q-Fly: An Optical Interconnect for Modular Quantum Multicomputers}

\author{Daisuke Sakuma}
\authornote{Both authors contributed equally to this research.}
\email{sakumadaisuke32@gmail.com}
\orcid{0000-0002-1576-2185}
\affiliation{%
  \institution{Keio University}
  \city{Fujisawa}
  \state{Kanagaya}
  \country{Japan}
}

\author{Tomoki Tsuno}
\authornotemark[1]
\email{tsuno-tomoki-ks@ynu.jp}
\orcid{1234-5678-9012}
\affiliation{%
  \institution{Yokohama National University}
  \city{Yokohama}
  \state{Kanagawa}
  \country{Japan}
}

\author{Hikaru Shimizu}
\authornotemark[1]
\email{shimihika2357@keio.jp}
\orcid{1234-5678-9012}
\affiliation{%
  \institution{Keio University}
  \city{Yokohama}
  \state{Kanagawa}
  \country{Japan}
}

\author{Yuki Kurosawa}
\email{kyuryu@sfc.wide.ad.jp}
\orcid{1234-5678-9012}
\affiliation{%
  \institution{Keio University}
  \city{Fujisawa}
  \state{Kanagawa}
  \country{Japan}
}

\author{Monet Tokuyama Friedrich}
\email{bob@sfc.wide.ad.jp}
\orcid{1234-5678-9012}
\affiliation{%
  \institution{Keio University}
  \city{Fujisawa}
  \state{Kanagawa}
  \country{Japan}
}

\author{Kentaro Teramoto}
\email{zigen@mercari.com}
\orcid{1234-5678-9012}
\affiliation{%
  \institution{Mercari, Inc.}
  \city{Minato}
  \state{Tokyo}
  \country{Japan}
}

\author{Amin Taherkhani}
\email{amin@sfc.wide.ad.jp}
\orcid{1234-5678-9012}
\affiliation{%
  \institution{Keio University}
  \city{Fujisawa}
  \state{Kanagawa}
  \country{Japan}
}

\author{Andrew Todd}
\email{at@auspicacious.org}
\orcid{1234-5678-9012}
\affiliation{%
  \institution{Keio University}
  \city{Fujisawa}
  \state{Kanagawa}
  \country{Japan}
}

\author{Yosuke Ueno}
\email{yosuke.ueno@riken.jp}
\orcid{1234-5678-9012}
\affiliation{%
  \institution{RIKEN}
  \city{Wako}
  \state{Saitama}
  \country{Japan}
}

\author{Michal Hajdu\v{s}ek}
\email{michal@sfc.wide.ad.jp}
\orcid{1234-5678-9012}
\affiliation{%
  \institution{Keio University}
  \city{Fujisawa}
  \state{Kanagawa}
  \country{Japan}
}

\author{Rodney Van Meter}
\email{rdv@sfc.wide.ad.jp}
\orcid{1234-5678-9012}
\affiliation{%
  \institution{Keio University}
  \city{Fujisawa}
  \state{Kanagawa}
  \country{Japan}
}

\author{Rikizo Ikuta}
\email{ikuta.rikizo.es@osaka-u.ac.jp}
\orcid{1234-5678-9012}
\affiliation{%
  \institution{The University of Osaka}
  \city{Toyonaka}
  \state{Osaka}
  \country{Japan}
}

\author{Toshihiko Sasaki}
\orcid{1234-5678-9012}
\affiliation{%
\institution{The University of Tokyo}
  \city{Tokyo}
  \country{Japan}
}
\authornote{Current address: Quantinuum Inc.}
\email{sasaki@qi.t.u-tokyo.ac.jp}

\author{Shota Nagayama}
\email{shota@qitf.org}
\orcid{1234-5678-9012}
\affiliation{%
  \institution{Keio University}
  \city{Yokohama}
  \state{Kanagawa}
  \country{Japan}
}


\begin{abstract}
Much like classical supercomputers, scaling up quantum computers requires an optical interconnect.
However, signal attenuation leads to irreversible qubit loss, making quantum interconnect design guidelines and metrics different from conventional computing.
Inspired by the classical Dragonfly topology, we propose a multi-group structure where the group switch routes photons emitted by computational end nodes to the group's shared pool of Bell state analyzers (which conduct the entanglement swapping that creates end-to-end entanglement) or across a low-diameter path to another group. We present a full-stack analysis of system performance, a combination of distributed and centralized protocols, and a resource scheduler that plans qubit placement and communications for large-scale, fault-tolerant systems. We implement a prototype three-node switched interconnect to justify hardware-side scalability and to expose low-level architectural challenges. We create two-hop entanglement with fidelities of 0.6-0.76. Our design emphasizes reducing network hops and optical components to simplify system stabilization while flexibly adjusting optical path lengths. Based on evaluated loss and infidelity budgets, we find that moderate-radix switches enable systems meeting expected near-term needs, and large systems are feasible. Our design is expected to be effective for a variety of quantum computing technologies, including ion traps and neutral atoms.
\end{abstract}

\maketitle

\section{Introduction}

\begin{figure}
    \centering
    \includegraphics[width=1\linewidth]{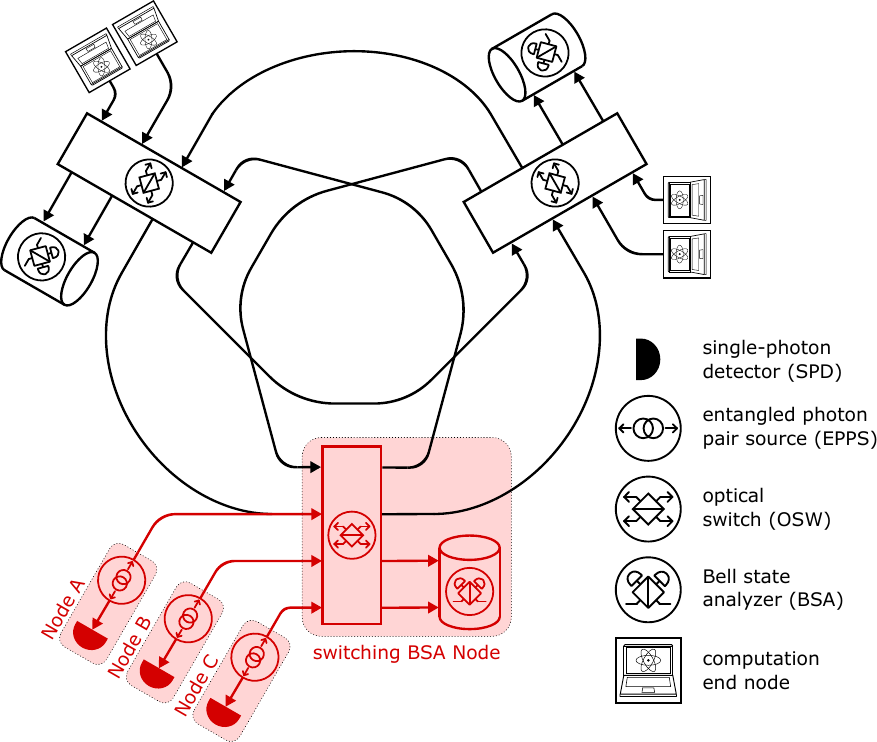}
    \Description[Three nodes and switches arranged in a triskele.]{Three nodes and switches arranged in a triskele.  Each switch has an arrow connecting it to the next group counter-clockwise around the loop.  Each switch has two computational node connected to it.  One group is in red with 3 end nodes (the additional node is considered as the alternative input to the red group switch).}
    \caption{A three-group Q-Fly with the experimental demonstration shown in red. Entangled pairs of photons are generated at the EPPS nodes. One photon of each pair is measured by the single-photon detectors (SPDs). The remaining photons are guided to the switching BSA node, composed of an optical switch (OSW) and a Bell state analyzer (BSA). The BSA and other detectors are housed inside a cryostat, represented by the cylinder.}
    \label{fig:q-fly-experiment}
\end{figure}


Recognition is growing that extending beyond advances in monolithic systems~\cite{google2019quantum, IBM, MuraliISCA19, MinISCA23}, modular architectures are the critical pathway to truly scalable quantum computers~\cite{van-meter16:_ieee-comp,PRXQuantum.2.017002,IBM02}.  Single devices, such as a single superconducting chip or an ion trap, are limited to at most a few thousand physical qubits, but fault-tolerant systems of sufficient scale to solve large problems will require huge numbers of qubits, with proposed systems ranging well into the millions.  One proposed direction to achieve this level of scale is for quantum nodes to hold one or more logical qubits, and perform logical gates remotely or by teleporting data~\cite{oi06:_dist-ion-trap-qec,van-meter06:thesis,nagayama17:thesis,Ahsan:2015:DMQ:2856147.2830570,guinn2023codesignedsuperconductingarchitecturelattice,leone2024resourceoverheadsattainablerates, IsailovicISCA06}.  Other proposals rely on error correction code words that span multiple nodes~\cite{jiang07:PhysRevA.76.062323,van-meter10:dist_arch_ijqi,nickerson2013topological,nickerson14:PhysRevX.4.041041,de-bone24:dist-qec-thresh}. Recent work has also studied the possibility of noisy intermediate-scale quantum (NISQ) multicomputers~\cite{ang:10.1145/3674151,Preskill2018}.
Regardless of the approach taken towards scalability, interconnects are widely recognized as being crucial not only for performance but also for correctness and robustness of computation.

The challenge is to demonstrate that the physical system will scale up in number of end nodes as we increase the size and complexity of the switching network, which will negatively affect the key metrics of \emph{photon loss} (measured in decibels) and \emph{fidelity}, a measure of the quality of a quantum state; roughly, the probability that the system currently holds the state it should hold.
Quantum interconnect architectures can learn from classical \emph{multicomputer} architectures, which are the standard approach to supercomputing systems today~\cite{athas:multicomputer,dally04:_interconnects,ajima:10.1109/MC.2009.370,kim2008dragonfly-technology,shpiner17:dragonfly-plus,stunkel20:summit-sierra,de-sensi20:slingshot,besta2014slimfly}.
However, a number of key characteristics of quantum systems mean that this application is far from straightforward.
We highlight these fundamental challenges below.





\emph{Every decibel matters.}
In classical systems with optical-electrical-optical switches or purely electrical switches, signals are regenerated at each switch.
For quantum switches working at the level of single photons, this is fundamentally prohibited by the \emph{no-cloning theorem}~\cite{Wootters1982no-cloning}.

As a hypothetical example, if the loss in an optical path increases by 6dB, the performance will fall by a factor of four and so the memory lifetime demands will increase by a factor of four.  This in turn might require us to raise the error correcting code distance, say from 9 to 11.  Using the surface code~\cite{Fowler2008a1}, this will result in $\sim (11/9)^2 \approx 1.5$ times the physical qubits and 22$\%$ increase in the time needed for each logical gate, including the remote gates, resulting in $5\times$ slower performance at $1.5\times$ the hardware cost.

Overcoming the loss of individual photons requires complex nodes, known as \emph{quantum repeaters}~\cite{Briegel1998}, which are necessary in wide-area systems but which we would like to avoid in data center-scale systems.
Instead, these systems are expected to rely on purely optical end-to-end paths.
Thus, while network diameter influences performance, the loss across the group switch itself matters and, for some switch technologies, is strongly group size-dependent. These factors lead to the end-to-end decibel-oriented mantra above, rather than the ``It's ALL about the diameter,'' mantra of Slimfly~\cite{besta2014slimfly}.
    

\emph{Error rates are very high.} Even when received, the error rates for single photons, taking into account generation, propagation, switching, and measurement, typically amount to a few percent.  This is far too high for correct computation, so entanglement generation will be followed by error detection in the form of \emph{purification}~\cite{dur2007epa} or error correction~\cite{Devitt2013qecforbeginners}.
    


\emph{Operations are probabilistic.} Another major impact on system performance is that some crucial operations are inherently probabilistic.
A notable example is \emph{optical entanglement swapping}, also known as \emph{Bell state analysis}, where the success probability is fundamentally limited to 50\% when implemented with linear optics.
Failure of such probabilistic operations necessitates retrial at a significant performance cost.

\emph{Synchronization demands can be very strict.} The length of a photon's wave packet can range over a startlingly large six or more (decimal) orders of magnitude, depending on the chosen technology.
In some cases, including our experiments, they can be as short as a couple of millimeters, which corresponds to only a few picoseconds.  Since some tasks require that photons arrive simultaneously, our synchronization system must be tuned to a small fraction of that time.
    
\emph{The flow of photons is unidirectional,} but applications can move data in either direction.  Once the interconnect establishes entanglement between two end nodes, \emph{quantum teleportation}~\cite{Bennett1993} is used to move qubits from one node to another, or the entanglement is used to execute a \emph{remote quantum gate}~\cite{eisert2000oli,van-meter07:_distr_arith_jetc}.  Once created, the behavior of the entanglement is independent of the directions in which the photons originally propagated.
    
\emph{Good detectors are expensive and require a cryogenic environment.} Room temperature detectors are common but moderate in loss; given that multiple detectors must fire simultaneously, the multiplicative loss of performance is unacceptable. Current state of the art detectors are needed for each link, but these require a cryogenic environment, complicating physical layout.  Moreover, fabrication yields are low and supporting electronics are complex.
    
    
    

\emph{Reconfiguring connections is slow and recalibration must be performed frequently.} Reconfiguring the switches themselves is not the bottleneck in the process; in order to tune each new connection and ensure optimal operation, additional components (currently mechanical) must be adjusted.
Such adjustments can take as long as a few seconds~\cite{krutyanskiy2023entanglement}, resulting in a significant decrease in overall system performance.
Furthermore, due to the extreme sensitivity of single photons to environmental noise, the system must be frequently recalibrated, even if no new connections need to be accommodated.




Given these unique challenges, the current focus of modular quantum system research is the link~\cite{alshowkan2022advanced,pompili2022experimental} and connection layers~\cite{rao2023throughput,alshowkan2021reconfigurable}.
This is a necessary first step towards building more complex systems.
However, in order to bring us closer to scalable and robust distributed quantum systems, there is a pressing need for deeper understanding of higher-layer concepts such as switching network architectures, and their impact on the performance when executing large-scale quantum computation.

We address this need by presenting a full-stack architecture for a quantum switching network.
Our approach is guided by the following principles:

\begin{itemize}
    \item In order to curtail the effect of photon loss, the switching interconnect architecture should minimize the number of hops and optical elements, while allowing connectivity between all pairs of end nodes.
    \item The proposed architecture should be implementable with present-day hardware, and allow for fully reconfigurable and robust control.
    \item The switching architecture must allow for efficient execution of distributed quantum computational primitives.
    \item Finally, the architecture should provide a clear path for technological evolution.
\end{itemize}

Our contributions are summarized below.
\begin{itemize}
    \item We present a scalable switching architecture, referred to as ``Q-Fly'' (Fig.~\ref{fig:q-fly-experiment}), based on the group and inter-group structure of the Dragonfly~\cite{kim2008dragonfly-technology}. This architecture avoids the higher hop counts of other network architectures such as fat trees~\cite{leiserson85:fat-tree,leiserson1992network}.
    We introduce three Q-Fly topologies, with differing inter-group connectivities and optical path depths, and perform end-to-end loss analysis.
    \item We experimentally implement a local group of the Q-Fly, given by three measurement end nodes, distributing entangled photon pairs via a switchable Bell state analyzer (BSA), as highlighted in red in Fig.~\ref{fig:q-fly-experiment}.
    Using bulk optical elements and an automated, reconfigurable control system, we demonstrate excellent stability of the system, producing post-switched (two-hop) end-to-end fidelities in the range of 0.60-0.76 at an average rate of 5 Hz.
    \item We experimentally demonstrate the scalability of the Q-Fly group switch.
    \item We project the performance of the Q-Fly architecture in a large-scale quantum computation by creating execution schedules for a distributed version of the quantum Fourier transform (QFT).
    We identify the parameter regimes where the Q-Fly topology performs more than 2 times faster than a 2D lattice, and only 2.1 times slower than a monolithic execution.
\end{itemize}

We begin by outlining the fundamental quantum technologies required for an optical switching network in Sec~\ref{sec:primitives}.
Using these quantum primitives, we define three Q-Fly variants in Sec.~\ref{sec:q-fly}, along with the quantum optical system implementing a group switch. 
We evaluate the performance of our experimental implementation in Sec.~\ref{sec:eval}.
In order to study the projected performance of Q-Fly in the context of large-scale fault-tolerant quantum computation in Sec.~\ref{sec:projected}, we develop a qubit allocation procedure and network scheduler along with a distributed quantum computation performance evaluation tool.
Finally, we discuss the set of conditions under which Q-Fly is a viable architectural choice in Sec.~\ref{sec:discuss}.

\section{Quantum Network Primitives}
\label{sec:primitives}




Entanglement is a correlation shared between two or more quantum subsystems that has no parallel in classical systems~\cite{horodecki2009quantum}.
It is a fundamental resource that is exploited in the vast majority of quantum technologies, such as secure communication~\cite{ekert1991quantum}, distributed quantum computing~\cite{barral2024review,caleffi2024distributed}, and sensing~\cite{gottesman2012longer}.

The primary function of quantum networks is to distribute entanglement between distant nodes in order to enable these quantum applications.
This is typically done with the help of pairs of photons entangled in their polarization degree of freedom.
One such entangled state of photons $A$ and $B$ is denoted by $\ket{\psi^+}\u{AB}=(\ket{HV}\u{AB}+\ket{VH}\u{AB})/\sqrt{2}$, where $\ket{H/V}$ denotes the horizontally/vertically polarized state of the photon.
In a quantum network, these states are generated by the \emph{entangled photon pair source} (EPPS).
In reality, these states will be affected by various sources of noise, resulting in deviations from the ideal state $\ket{\psi^+}$.
Such noisy states are denoted by the \emph{density operator} $\rho_{AB}$, and represent our lack of knowledge about the real distributed state.
Such deviations are quantified by \emph{fidelity}, denoted by $F$, which roughly represents the probability that the real distributed state is the ideal state $\ket{\psi^+}_{AB}$.
This is often represented as \emph{infidelity}, defined as $1-F$.



Entanglement generation between distant quantum systems that have not interacted previously is possible via \emph{entanglement swapping}.
Two separate entangled states $\ket{\psi^+}_{AB_1}$ and $\ket{\psi^+}_{B_2C}$ can be used to generate an entangled state $\ket{\psi^+}_{AC}$ by performing a \emph{Bell measurement} on qubits $B_1$ and $B_2$, and communicating the outcome classically to $A$ and $C$.
The quantum network node that performs this two-qubit measurement is referred to as a \emph{Bell state analyzer} (BSA).
It is crucial to note that BSA nodes implemented with linear optics have a maximum theoretical success probability of 0.5~\cite{vaidman1999methods}.



It is important to note that entanglement is not directional.
Quantum nodes $A$ and $B$ that share an entangled state $\ket{\psi^+}_{AB}$ can use this state to transfer quantum information from $A\rightarrow B$ or from $B\rightarrow A$ via \emph{quantum teleportation}~\cite{bennett1993teleporting}.
In the context of distributed quantum computation, nodes $A$ and $B$ can also use $\ket{\psi^+}_{AB}$ to execute a remote quantum gate~\cite{eisert2000oli}.

\section{The Q-Fly Topology and System Design}
\label{sec:q-fly}

\subsection{Network Topologies}
\label{sec:q-fly:topologies}

The Q-Fly, like other ``-fly'' topologies, achieves scalability by attaching computational (and possibly measurement, storage and sensing) end nodes to switches in \emph{groups}~\cite{kim2008dragonfly-technology,besta2014slimfly,shpiner17:dragonfly-plus,lakhotia22:polarfly}. Each group can consist of a set of switches that are strongly interconnected. The groups are connected to each other via one or more links. 

\begin{table}[]
    \centering
    \caption{Notation for Q-Fly architectures.}
    \begin{tabular}{c|p{6cm}}
         $N$ & number of end nodes in fully populated system \\
$g$ & number of groups in fully populated system \\
$p$ & group size (number of quasi-half duplex end nodes attached to one group switch) \\
$p'$ & group size (number of quasi-full duplex end nodes attached to one group switch) \\
$k$ & group switch radix (i.e., $k$ inputs and $k$ outputs) \\
$k_{\textrm{elm}}$ & when a group switch is build from smaller components, the radix of one element \\
$b$ & number of BSAs in one group (each 4 detectors, 2 switch egress ports) \\
$d$ & number of detectors in total system (BSAs only)
    \end{tabular}
    \label{tab:notation}
\end{table}

\begin{figure*}
    \centering
    \includegraphics[width=\textwidth]{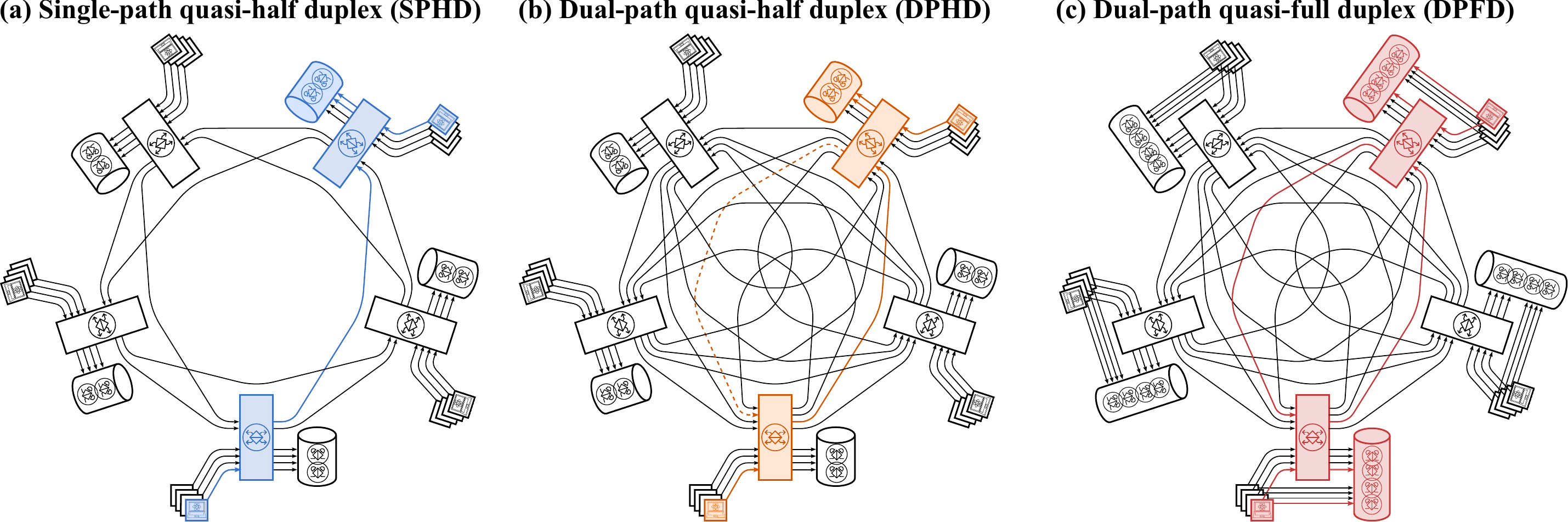}
    \Description[The figures are vertically illustrated in three sub-figures, each showing a different configuration of a Q-Fly with five switch groups arranged in a circular layout. Each group connects four quantum computational nodes to the left side of a switch, and a group of BSAs with varying counts is located on the right side of the switch. The interconnections between the switches differ in each sub-figure.]{In sub-figure (a) Each switch group has 6 input and 6 output ports. Four of the input ports are connected to computational nodes, while the remaining two come from two switches on the left. Among the six output ports, four are connected to BSAs, and the remaining two connect to two switches on the right. Sub-figure(b) is similar to (a),  but each switch group has 8 input and 8 output ports. Additionally, all switches are fully interconnected. The interconnections between switch groups in sub-figure (c) are similar to (b), but each computational node has an additional link directly connected to a specific BSA. As a result, there are four BSAs instead of two on the right side of each switch.}
    \caption{
    Three Q-Fly variants with an example connection between two end nodes located in separate groups. (a) Single-path quasi-half duplex (SPHD).
    (b) Dual-path quasi-half duplex (DPHD). Dashed orange line shows the second fiber connecting the two groups. 
    (c) Dual-path quasi-full duplex (DPFD). The two end nodes have two interfaces, allowing two paths to be used concurrently, and with fewer switches in the optical path.  
    }
    \label{fig:q-flies}
\end{figure*}

We develop quantum variants of the Dragonfly, where each group consists of a switch that supports full intra-group communication, and some ports of the switches are used for inter-group communication.  Due to the need to route photons to BSAs, the architecture has only two tiers of connections: intra-group and inter-group (in contrast to the three-tier structure of Dragonfly).
We introduce three Q-Fly architectures, as shown in Fig.~\ref{fig:q-flies}. Our notation is listed in Tab.~\ref{tab:notation} and the construction parameters are compared in Tab.~\ref{tab:topology-comparison}.
In general, classical fly designs are drawn as undirected graphs and it is understood that each line represents a full duplex link, e.g. a pair of counter-propagating fibers.  Here, we will always be explicit about photon propagation direction and lines in the figures are assumed to be single-direction channels.

\emph{Single-path, quasi-half duplex} topologies are constructed by placing the $g$ groups in a circle, as shown in Fig.~\ref{fig:q-flies}(a).  Each of the $p$ end nodes in a group has one unidirectional connection from the node to a group switch.  The outputs of the switch are divided into two sets; $2b = p$ ports to the $b$ BSAs in the BSA pool and $\lfloor g/2\rfloor$ ports to other groups. The group switch radix is $k = p+\lfloor g/2\rfloor$. Each group adds a fiber from itself to the $\lfloor g/2\rfloor$ groups on its right, going counterclockwise.
Note that each end node has only a single interface, though the established entanglement can be used to communicate in either direction.
Therefore, we refer to this case as \emph{quasi-half duplex} MIM link.
This architecture is the simplest to build, but has the lowest inter-group bandwidth and provides no fault tolerance for single-hop group-to-group connections.

\emph{Dual-path, quasi-half duplex} topology is shown in Fig.~\ref{fig:q-flies}(b).  Each group adds a fiber from itself to every group on its right, going counterclockwise, continuing all the way around the system.  Each of the $p$ end nodes in a group has one unidirectional connection from the node to a group switch.  The outputs of the switch are divided into two sets; $2b = p$ ports to the $b$ BSAs in the BSA pool and $g-1$ ports to other groups. The group switch radix is $k = p+g-1$.
Note that a fiber carries single photons from every group to every other group.
This means that each pair of groups is connected by two fibers running in opposite physical directions, providing bandwidth and fault tolerance.
However, each end node has only a single interface, making the connections quasi-half duplex.

\emph{Dual-path, quasi-full duplex} topologies are depicted in Fig.~\ref{fig:q-flies}(c).
Unlike previous cases, the end nodes have two interfaces.
One interface of each of the $p'$ end nodes is connected directly to one port of a currently unconnected group BSA, and the second port is connected to an input port of the group switch. Next, one output port from the group switch is connected to the second port of each of the $b = p'$ BSAs. The remaining group switch ports are dedicated to inter-group communication, following the same procedure as the dual-path, quasi-half duplex topology. Note that the number of nodes supported for a given radix is the same as the dual-path, quasi-half duplex configurations, but twice as many BSAs and detectors are needed.

\begin{table}[]
\centering
\caption{Comparison of system scaling characteristics for single-path, quasi-half duplex (SPHD), dual-path, quasi-half duplex (DPHD), and dual-path, quasi-full-duplex (DPFD) topologies. These configurations maximize total nodes at the expense of group-to-group bandwidth and redundancy.}
\begin{tabular}{lrrrrrrl}
\hline
 Type & $k$ &       $N$ & $g$ & $p$ or $p'$  & $b$ &     $d$ \\
\hline
 SPHD & 6 & 20 & 5 & 4 & 2 & 40 \\
 DPHD & 6 & 12 & 3 & 4 & 2 & 24 \\
 DPFD & 6 & 12 & 3 & 4 & 4 & 48 \\
 \hline
 SPHD & 8 & 36 & 9 & 4 & 2 & 72 \\
 DPHD & 8 & 20 & 5 & 4 & 2 & 40 \\
 DPFD & 8 & 20 & 5 & 4 & 4 & 80 \\
 \hline
 SPHD & 16 & 136 & 17 & 8 & 4 & 272 \\
 DPHD & 16 & 72 & 9 & 8 & 4 & 144 \\
 DPFD & 16 & 72 & 9 & 8 & 8 & 288 \\
 \hline
 SPHD & 576 & 166176 & 577 & 288 & 144 & 332352 \\
 DPHD & 576 & 83232 & 289 & 288 & 144 & 166464 \\
 DPFD & 576 & 83232 & 289 & 288 & 288 & 332928 \\
 \hline
 SPHD & 1100 & 605550 & 1101 & 550 & 275 & 1211100 \\
 DPHD & 1100 & 303050 & 551 & 550 & 275 & 606100 \\
 DPFD & 1100 & 303050 & 551 & 550 & 550 & 1212200 \\
\hline
\end{tabular}
\label{tab:topology-comparison}
\end{table}


\subsection{Construction of a Group Switch}

When constructing the group switch, it is essential to analyze the performance and loss of currently available full-optical switching technologies. Different choices of switch architecture, topology, and switching mechanism have a significant impact on performance indicators such as insertion loss.
We consider three types of components for the construction of a full radix-$k$ group switch: discrete 2$\times$2 switching elements, planar photonic chips, and $k_{\textrm{elm}}\times k_{\textrm{elm}}$ switches where $k_{\textrm{elm}}$ denotes the number of ports in an individual switching element.

We denote the fiber-to-BSA insertion loss of a $k$-radix group switch as GroupLoss($k$).
When the group switch is constructed from discrete $2\times 2$ switches, we can use a Bene\v{s} network, which requires a switch depth of $2\lceil \log_2{k}\rceil -1$. We let $x_{\text{2$\times$2}}$ [dB] be the fiber-to-fiber insertion loss of a $2\times 2$ switch. In this setting, GroupLoss($k$) becomes $(2\lceil \log_2{k}\rceil -1) x_{\text{2$\times$2}}$ [dB].

For planar photonic chips, which are restricted to a planar arrangement of $2\times 2$ switch cells, we can use a planar permutation network~\cite{Spanke:87}, requiring a switch depth of $k$. Given $x_{\text{sp-coupling}}$ [dB] and $x_{\text{sp-2$\times$2}}$ [dB] as the coupling loss and the $2\times 2$ switching cell loss respectively, GroupLoss($k$) becomes $x_{\text{sp-coupling}} + k x_{\text{sp-2$\times$2}}$ [dB].

If the group switch is constructed from a single $k\times k$ switch with $x_{\text{$k\times k$}}$ [dB] insertion loss, GroupLoss($k$) is simply $x_{\text{$k\times k$}}$ [dB].


\subsection{Choosing the Configuration}

Configuring a network is a large combinatoric space with many pareto optimal points, depending on the key metrics chosen to optimize.  For Q-Fly, our goal is to minimize execution time for a specified workload, assuming that availability of detectors is a key constraint.  We would like to explore the whole combinatoric space, but in lieu of that exhaustive process, we present a heuristic approach to matching the system configuration to the workload.

We assume that the early part of the system design has been done; the physical node technology is understood, a code and code distance have been selected, and we know the minimum number of end nodes $N_{\textrm{min}}$ that will be needed.  Further, we have considered architectural options and settled on quasi-half duplex Q-Fly.  Our configuration problem then is to select the number of groups $g$, the group switch radix $k$, and the group size $p$.

Begin by enumerating available switch candidates, as discussed in the previous subsection; the most important characteristics are the number of ports in an individual switch ($k_{\textrm{elm}}$) and the loss, though the switch's effect on polarization and switching time are also important. The loss of the group switch, especially a composite switch, is evaluated in more detail in Sec.~\ref{sec:2-node}.

We must select $g$ and $p$ such that $gp \ge N_{\textrm{min}}$.  If $N_{\textrm{min}} \le k_{\textrm{elm}}$,  then $g=1, p=N_{\textrm{min}}$  representing a single-switch, single-group, fully connected crossbar; otherwise, a multi-group structure is required.  In the multi-group structure, balancing the number of groups $g$ with the group size $p$ is traffic dependent.

If $N_{\textrm{min}} \leq k_{\textrm{elm}}^2 / 4$, each group switch can be exactly one physical switching element.  For $N_{\textrm{min}} \approx k_{\textrm{elm}}^2 / 4$, select $g \approx p \approx k_{\textrm{elm}} / 2$.  For $N_{\textrm{min}} \ll k_{\textrm{elm}}^2 / 4$, choose small $g$, large $p$ with single inter-group links if traffic has high locality.  If instead the traffic has low locality, choose small $p$, large $g$ with multiple inter-group links.  For $k_{\textrm{elm}} < \sqrt{N_{\textrm{min}}}/2$, each group switch will be a composite switch comprising several switching elements.  Large group sizes $p$ will result in a composite group switch with higher depth and higher loss, slowing down intra-group traffic.  Therefore, minimizing intra-group switch loss will happen for $p ~= g ~= \sqrt{N_{\textrm{min}}} / 2$.

Following this heuristic and the system construction process outlined in Sec.~\ref{sec:q-fly:topologies} will give a system with reasonable balance that can be further tuned later.  This analysis can be repeated using several choices for the switch technology and group switch construction.

\section{Prototype Implementation}
\label{sec:eval}

We demonstrated the feasibility of our architecture by constructing the red portion of Fig.~\ref{fig:q-fly-experiment}. This corresponds to entanglement distribution between two nodes selected by a network protocol, through a group switch. It reveals the actual procedure of the entanglement distribution and provides insight into performance estimation. The prototype implementation begins with a baseline experimental setup described in Section~\ref{sec:DesignPrototype}, followed by data and analysis in the next two subsection. To demonstrate the scalability of the design, we then extend this setup to emulate a larger group switch.
 
\subsection{Design of the Prototype}\label{sec:DesignPrototype}


\begin{figure*}[t]
    \centering
    \includegraphics[width=1\linewidth]{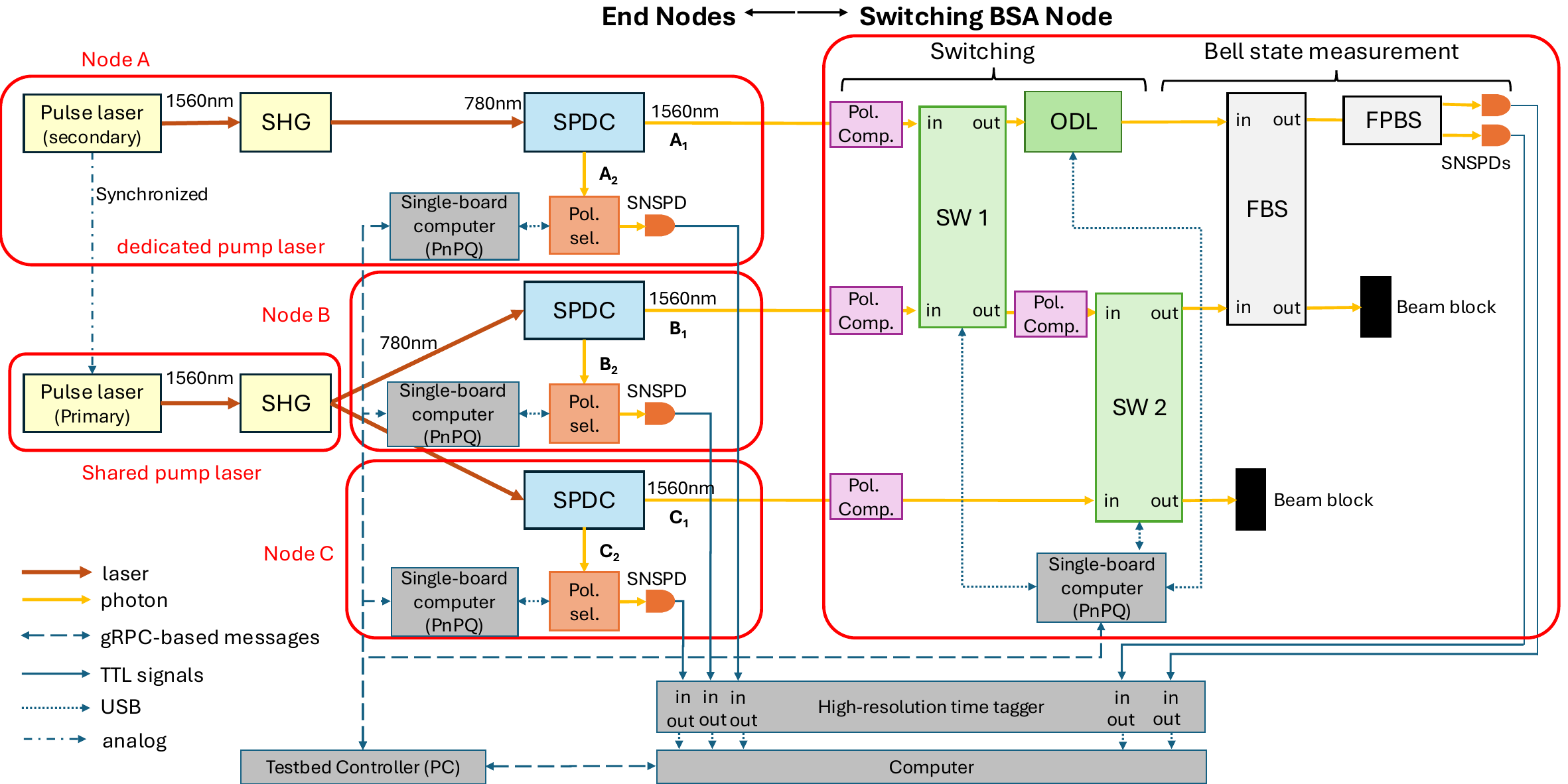}
    \Description[This diagram illustrates the overall setup for the experiment]{Boxes and arrows showing the flow of both photons and control signals, from the pump laser, through SHG and SPDC, through the switches, and to the measurement setups.  Control computers and signals are included.}
    \caption{Optical system, analog and digital control systems corresponding to the red portions of Fig.~\ref{fig:q-fly-experiment}. FBS, fiber beam splitter. FPBS, fiber polarizing   beam splitter.  ODL, optical delay line. Pol. comp., polarization compensator. Pol. sel., polarization selector. SHG, second harmonic generation (frequency doubler). SNSPD, superconducting nanowire single photon detector. SPDC, spontaneous parametric down conversion. SW, switch.} 
        \label{fig:optical-path}
\end{figure*}

The experimental setup is shown in Fig.~\ref{fig:optical-path}.
Each of the three end nodes~(A, B, C) can generate entangled photon pairs and measure one photon of the pair.
Switching BSA node links any pair of end nodes via entanglement swapping. 
At end node A, a mode-locked laser at \SI{1560}{nm}~(repetition rate: \SI{1}{GHz}, pulse width: \SI{5}{ps}, spectral width: \SI{0.7}{nm}) is amplified
by an erbium-doped fiber amplifier~(EDFA), and then frequency doubled by second harmonic generation~(SHG). 
The SHG light at \SI{780}{nm} pumps the degenerate photon pair at \SI{1560}{nm} by spontaneous parametric down conversion~(SPDC) in a type-II PPLN waveguide. 
By separating the photon pair into two different spatial paths using a half beamsplitter~(HBS),
the polarization-encoded entangled photon pair
$\ket{\psi^+}\u{A_1A_2}=(\ket{HV}\u{A_1A_2}+\ket{VH}\u{A_1A_2})/\sqrt{2}$
is prepared by postselection with probability of 0.5.

The photon in mode (path) $A_1$ is coupled to a single-mode fiber~(SMF) and sent to the switching BSA node. 
The photon in mode $A_2$ passes through a polarization selector~(Pol. sel.) composed of 
a quarter-wave plate~(QWP), a half-wave plate~(HWP), and a polarizing beamsplitter~(PBS). 
Then, it is coupled to an SMF and sent to the detector for node A. 

At nodes B and C, another mode-locked pulse laser is used. 
The repetition rate is \SI{1}{GHz}, which is determined by the internal clock of the laser device. 
The \SI{1}{GHz} clock signal is connected to the laser source used in node A to synchronize their repetition rates. 
After the SHG, the \SI{780}{nm} pulse is coupled to a polarization-maintaining fiber 
and distributed as the pump lights to prepare the SPDC photon pairs 
$\ket{\psi^+}\u{B_1B_2}$ and $\ket{\psi^+}\u{C_1C_2}$ at nodes B and C, respectively.  
Like node A, the photons in modes $B_1$ and $C_1$ are sent to the switching BSA node, 
and the photons in modes $B_2$ and $C_2$ are detected locally. 

At the switching BSA node, two of the three photons $A_1$, $B_1$ and $C_1$ are selected 
using two optical switches~(SW 1 and SW 2) as inputs to the BSA, which is 
composed of a fiber-based beamsplitter~(FBS), a fiber-based polarizing beam splitter~(FPBS), and two single-photon detectors. 
The BSA succeeds when the photons are detected at the two different detectors, which occurs with probability of $1/4$.
This transforms the two-photon states at the end nodes $x$ and $y$~($x,y=A,B,C$) from 
$\ket{\psi^+}_{x1x2}\ket{\psi^+}_{y1y2} \rightarrow \ket{\psi^+}_{x2y2}$.
For a successful entanglement swap, 
the polarization of each photon is adjusted by a polarization compensator. 
The arrival times of the photons are adjusted using an optical delay line~(ODL). 

Each photon is spectrally narrowed by a frequency filter with a bandwidth of \SI{0.28}{nm}. 
All of the single-photon detectors used in this experiment are 
superconducting nanowire single-photon detectors~(SNSPDs). 
Each of the SNSPDs has a time resolution of \SI{35}{ps} and a quantum efficiency of \SI{90}{\%}. 
The electrical signals coming from SNSPDs 
are collected by a multi-channel time tagger 
with a time resolution of \SI{5}{ps}. 

\subsection{Evaluation of the Prototype}
\begin{figure}[t]
    \centering
    \includegraphics[width=\columnwidth]{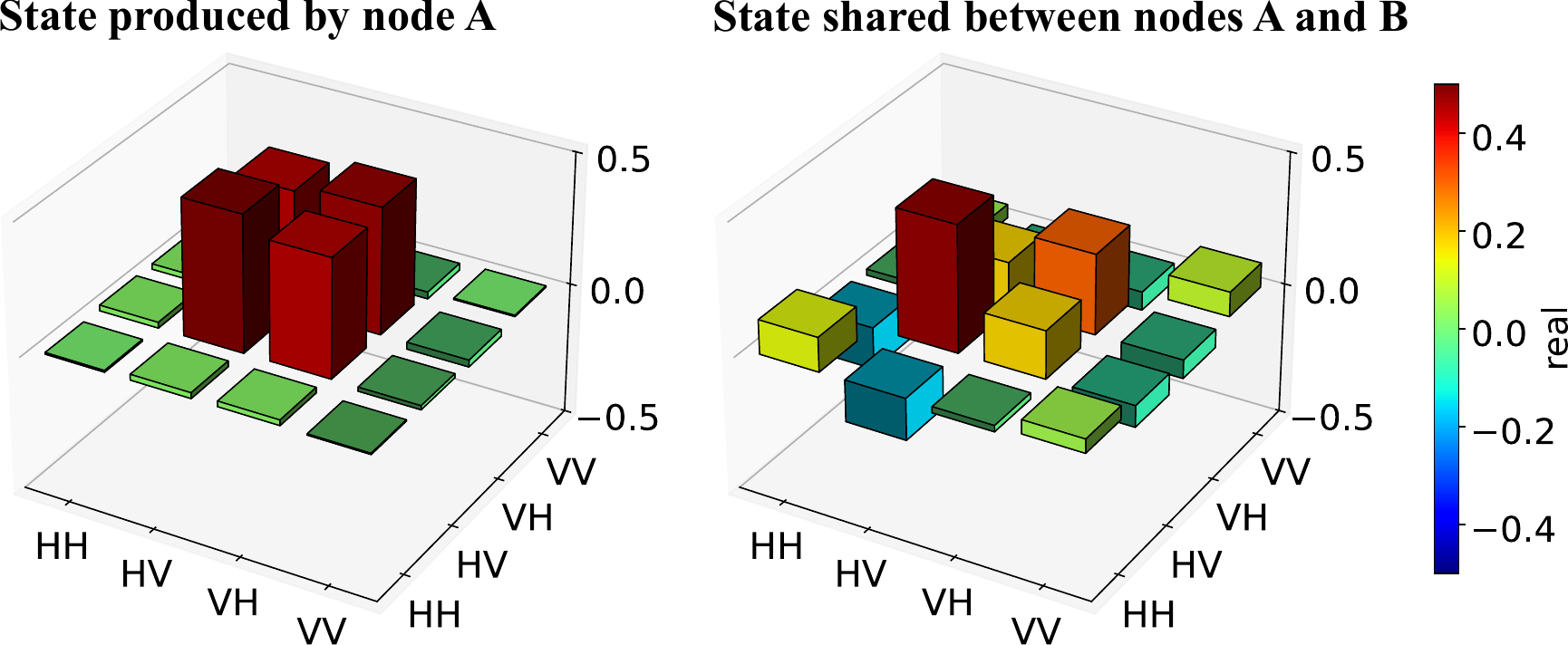}
    \Description[The diagrams shows a 3D representation of the real part of the density matrix related to node A]{}
    \caption{Real parts of
    $\rho_{A_1A_2}$ (left) and $\rho_{A_2B_2}$  (right). While fidelity has declined, entanglement sufficient for a variety of purposes is still present.
    }
    \label{fig:Density matrix}
\end{figure}

We evaluated the SPDC photon pairs produced at each node $x$~($x=A,B,C$) by installing the polarization analyzer into the optical path of photon $x_1$ 
and directly connecting the SMF to the SNSPD.
We performed quantum state tomography on photons in modes $x_1$ and $x_2$. 
Using the iterative maximum likelihood method, we reconstructed the density operator $\rho$~\cite{PhysRevA.64.052312}.
The result for node A is shown in Fig.~\ref{fig:Density matrix}. 
The fidelities of the photon pairs prepared at nodes A, B and C
are $0.94\pm 0.02$, $0.94\pm 0.01$ and $0.88\pm 0.01$, respectively. 
These results show that our nodes successfully prepare highly entangled states. 

\begin{figure}[t]
    \centering
    \includegraphics[width=0.45
    \textwidth]{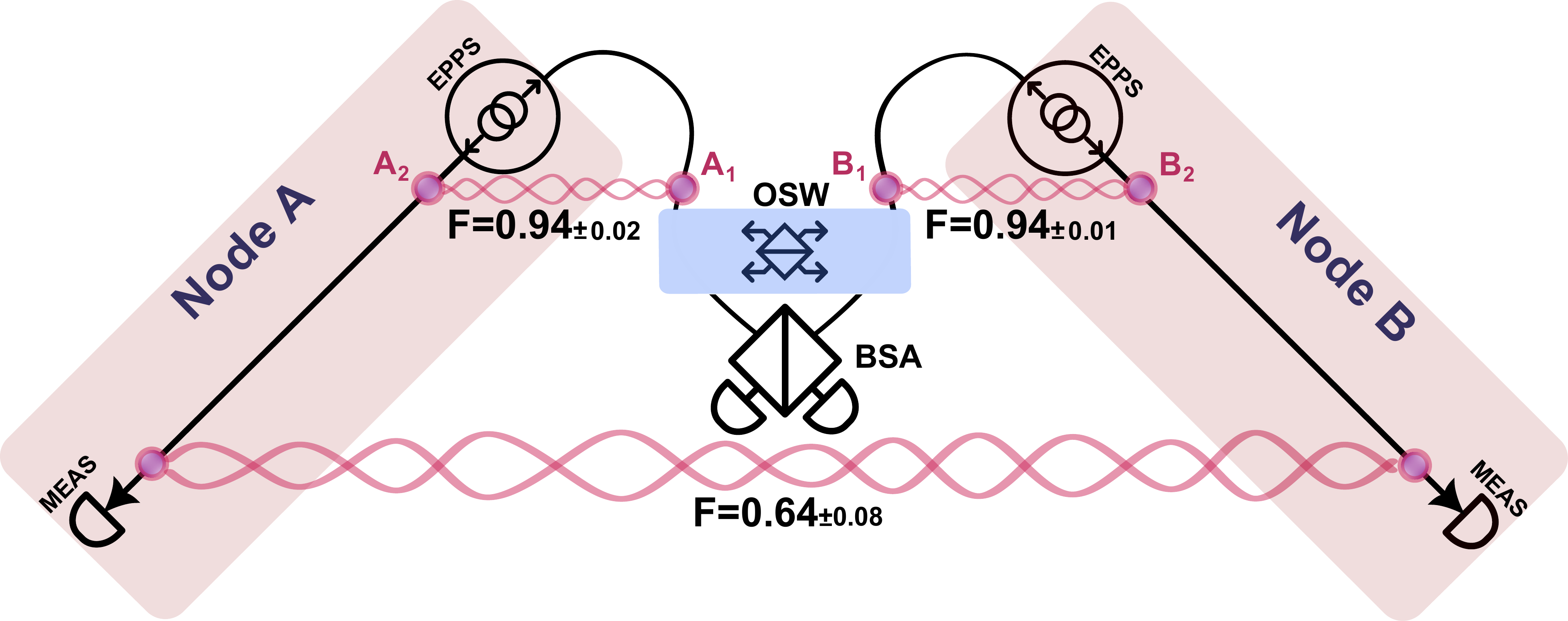}
    \Description[]{}
    \caption{Entanglement swapping performed on photons $A_1$ and $B_1$ after passing through the optical switch (OSW), leading to entanglement between photons $A_2$ and $B_2$.}
    \label{fig:Entanglement_swapping_diagram}
\end{figure}

Finally, we performed entanglement swapping as shown in Fig.~\ref{fig:Entanglement_swapping_diagram}. 
First, we set the position of the ODL to maximize the Hong-Ou-Mandel (HOM) visibility (see App.~\ref{sec:tuning-appendix}). 
We reconstructed the density operators of the two-photon states $(A, B)$, $(B, C)$ and $(C, A)$. 
The result for pair $(A, B)$ is shown in Fig.~\ref{fig:Density matrix}.
The fidelities of the states are 
$0.64\pm 0.08$, $0.61\pm 0.06$, and $0.60\pm 0.07$, respectively.
The values are well above 0.5, the minimum threshold for a pair of qubits to be entangled.
Thus, we conclude that entanglement swapping with the switching operation was successfully demonstrated.


\subsection{Evaluation of Loss and Infidelity}
 
We list the loss caused by each of the optical components in our experiment in Tab.~\ref{tab:lossbudget}. 
\begin{table}[]
    \centering
    \caption{Measured loss of optical circuits and components in the network, for the two photons of the entangled pairs generated at each node $x$~$(x = A, B, C)$. The quantum efficiency of each SNSPD is about \SI{90}{\%}~(equiv. to loss of \SI{0.45}{dB}). 
    Losses inside the nodes include \SI{3}{dB} loss of the HBS to split the SPDC photons. One photon in each pair ($x_2$) passes through one or two switches to the BSA and the ODL, as in the upper part of this table.  The other photon ($x_1$) is directly passed to the polarization selection and measurement, and is subject to only the loss in the last line.  Pho., photon.
    }
    \begin{tabular}{llr}
    \hline
Pho. & Component & Optical Loss [dB]\\ \hline 
$x_1$ & node $x$ inc. SNSPD & 14.7 (A), 14.3 (B), 14.8 (C) \\
&Optical switch~(SW 1) &  0.46 \\
&Optical switch~(SW 2) &  0.41 \\
&ODL     &  1.37 \\
&FBS      &  0.27 \\ 
&FPBS      &  0.46 \\ 
$x_2$ & node $x$ inc. SNSPD & 17.2 (A), 14.5 (B), 14.5 (C) \\ \hline 
    \end{tabular}
    \label{tab:lossbudget}
\end{table}
The infidelity after the entanglement swapping is mainly caused by the imperfection of the HOM visibility, caused by synchronization and polarization mismatch at the BSA.
From the experimental parameters, the impacts of the bandwidth and the multi-photon effects of the SPDC process on the HOM visibility are both estimated to be 0.9.
The effects of other imperfections on the HOM visibility, such as polarization mismatch or fluctuations during the experiment 
and the imbalance between the transmittance and reflectance of the FBS, 
are 0.9 for node pair $(A, B)$, and 0.8 for node pairs $(B, C)$ and $(C, A)$. 
The effects of imperfections outside of the SPDC process are 0.96, 0.84 and 0.82 for node pairs $(A, B)$, $(B, C)$ and $(C, A)$.

The end-to-end fidelity is below $0.9^2=0.81$ that one would expect given the fidelities of the individual link-level entangled pairs. One cause of finite distinguishability is photon wavepacket overlap, or synchronization. Our system utilizes photons with a spread of only few picoseconds, making the photons challenging to overlap spatially. We expect that future integration of ion-trap memories, and the subsequent use of photons with much longer wavepackets, will improve the indistinguishability of the photons.

An additional concern (incorporated into the numbers above) is wavelength differences between the two separate pump lasers.  Because the $(B,C)$ swapped pair has similar fidelity to the $(A,B)$ and $(A,C)$ pairs, we conclude that our two pump lasers are sufficiently similar.  Thus, longer wavepackets and better spatial overlap appear to be the best route to higher post-swapping fidelity.

\subsection{Demonstrating Scalability}

To demonstrate scalability, which is a crucial system design requirement, we must show the effects of increasing the number of optical components on both fidelity and end-to-end entanglement generation. Three switching elements were connected in a looped configuration, allowing us to emulate up to six additional switching layers, in addition to the two switch devices used in the baseline experiment. As expected, adding each element added about 0.5dB loss; adding six $2\times 2$ switching elements to the optical path increased loss by 3.15dB, as calculated based on a coincidence count experiment.

A composite group switch built as a Bene\v{s} network of these components gives switch depth of $2\lceil \log_2{k}\rceil -1$.  A depth-9 group switch gives radix $k = 2^5 = 32$ and has a loss of about 4.5dB.  In the SPHD and DPHD networks (Fig.~\ref{fig:q-flies}(a) and (b)), with both photons passing through the group switch, intra-group performance will be about an order of magnitude lower than a direct connection, while in DPHD (Fig.~\ref{fig:q-flies}(c)), performance will drop by only a factor of three as one photon bypasses the switch.  Inter-group performance falls by an additional factor of three.

The effect on fidelity is more complex in our current network prototype.  For each path, polarization compensation is performed, and is currently hand-tuned.  We saw no switch number-dependent variation in these experiments, with fidelity ranging from $0.60\pm 0.07$ to $0.76\pm 0.10$ independent of switch depth.

Thus, we conclude that the scalability of our optical interconnect is sufficient for building large networks, even with the technology currently available in our lab; further technology selection and development of the group switch itself will improve scalability. Details of the experimental demonstration are in Appendix~\ref{sec:app-scalability}.
\section{Projected Performance of the Q-Fly}
\label{sec:projected}

Having measured the performance of one group switch equipped with three end nodes as well as confirmed scalability up to six switching layers, we can now evaluate the performance of production-scale systems.
Since we focus on the interconnect, we restrict ourselves to a single choice on each of the other major system parameters: physical technology, node design, quantum error correcting code, compilation method, and application.
Each computational node is designed for surface code error correction with lattice surgery~\cite{fowler:PhysRevA.86.032324,horsman2012surface,TanISCA24}.
Computation is organized and compiled according to the Game of Surface Codes approach~\cite{Litinski:Quantum.3.128,liu2023_qce_substrate}.
Each node is internally organized into a set of \emph{tiles} (which define the code distance), and one or more tiles is used as a \emph{patch} to hold a single logical qubit.
Some logical single-qubit gates must be decomposed into longer sequences of up to several hundred $T$ and $S$ gates, which in turn are executed on the surface code using gate teleportation~\cite{ross2016optimal}.
We adopt Gidney's state of the art $\ket{T}$ state cultivation, in which preparation of the ancilla state needed for non-Clifford gates costs about the same as a lattice surgery CNOT~\cite{gidney2024magicstatecultivationgrowing}~\footnote{It is worth emphasizing Gidney's observation that cultivation is the end point of a series of innovations that have reduced the execution cost of magic state preparation by a factor of a thousand.  This preparation will no longer be the performance bottleneck even in monolithic systems.}.  We choose a large-scale, $n$-qubit quantum Fourier transform (QFT) consisting of $O(n^2)$ gates as the workload, which is useful not only for Shor's factoring algorithm, the related quantum phase estimation problem, and some forms of arithmetic, but also for Jordan \emph{et al.}'s new decoded quantum interferometry (DQI)~\cite{jordan2024dqi,Shor1997,yoshioka2024hunting}.


Inter-node logical two-qubit gates are assumed to be conducted by creating a physical Bell pair, transforming each of the two physical qubits into a logical qubit, performing \emph{purification} (a form of error detection) using multiple logical Bell pairs, and finally using the high-fidelity Bell pair to perform lattice surgery-style logical gates~\cite{nagayama2016interoperability,horsman2012surface,Litinski:Quantum.3.128,eisert2000oli}.
We expect this approach to perform better than fully distributed lattice surgery~\cite{guinn2023codesignedsuperconductingarchitecturelattice,leone2024resourceoverheadsattainablerates}.

\subsection{Two-node Entanglement Generation}
\label{sec:2-node}

We assess loss and infidelity caused by Q-Fly when entanglement generation is performed between two end nodes.
Loss and infidelity depend mainly on the network topology of Q-Fly, the performance of the group switch, and the traffic pattern.

We assume that the traffic pattern is generated by a centralized compiler, along with the placement of variable qubits among the distributed nodes, resulting in a schedule of required logical Bell pairs.
The entanglement generation process follows these steps over many application-specific \emph{rounds}:

\begin{enumerate}
    \item The network scheduler determines what logical Bell pairs schedule can be generated simultaneously, and sets the group switches to their appropriate state.
    \item Entanglement generation attempts are executed until all logical Bell pairs from Step (2) are established.
    \item Current round is concluded, the network scheduler moves back to Step (1) and creates a new set of active connections.
\end{enumerate}


In order to estimate the duration of one round, we introduce the following notation.
The time required for the group switches to change state, including ODL adjustment, in Step (1) is denoted by $t_{\text{gsw}}$.
The time required to create a physical and logical Bell pair is denoted by $t_{\text{phys}}$ and $t_{\text{log}}$, respectively.
The duration of a single round is then given by $t_{\text{gsw}} + t_{\text{log}}^*$, where $t_{\text{log}}^*$ is the maximum over $t_{\text{log}}$ for all active logical Bell pairs in a round.

Time to generate a logical Bell pair $t_{\text{log}}$ between two nodes is $t_{\text{phys}}\cdot n_{\text{phys}}$, where $n_{\text{phys}}$ the number of physical Bell pairs required.
Three rounds of $X$ then $Z$ purification will consume at least $4^3 = 64$ physical Bell pairs.  While the details will vary depending on the exact density matrix and local gate fidelity, when the physical Bell pairs are above $90\%$ in fidelity, the success rate of the first round of purification will be above $80\%$ and successive rounds will have near unit success probability.  Thus, creating a logical Bell pair with infidelity around $10^{-9}$ will consume at least $64/0.8 = 80$ physical Bell pairs.
In order to account for the non-unit success probability of successive purification rounds, we assume $n_{\text{phys}}=100$ in our simulation, resulting in logical entanglement generation time of
$t_{\text{log}} = 100 t_{\text{phys}}$.




The critical criteria determining the performance of each individual optical path are loss and infidelity.  These characteristics are dependent on the number of group switches along both arms between the end nodes and the BSA, and on the loss of each group switch as a function of the group radix $k$, $\operatorname{GroupLoss}(k)$.

Paths between two end nodes can be classified into two categories: intra-group and inter-group.
Intra-group paths connect end nodes within a single group, while inter-group paths connect end nodes located in distinct groups.
From Fig.~\ref{fig:q-flies}, we observe that the minimum intra- and inter-path losses for SPHD variant of the Q-Fly is given by $2\times\operatorname{GroupLoss}(k)$ and $3\times\operatorname{GroupLoss}(k)$, respectively.
This is true for the DPHD variant as well.
On the other hand, the DPFD variant enjoys lower losses of $1\times\operatorname{GroupLoss}(k)$ and $2\times\operatorname{GroupLoss}(k)$, respectively, due the ability of one of photons to bypass the group switches and be routed immediately to the BSA.

The minimum path loss above can be directly generalized to paths that traverse an arbitrary number of group switches $N_{\text{gsw}}$.
For quasi-half duplex variants (SPHD and DPHD), the path loss is
\begin{equation}
    \operatorname{PathLoss}_{\mathrm{HD}} = 10 \log\frac{1}{2} + L \times l_f + (N_{\mathrm{gsw}} + 1) \mathrm{GroupLoss}(k).
    \label{formula:loss_Q-Fly_HD}
\end{equation}
For the the quasi-full duplex variant (DPFD), the path loss is lower since the factor of $N_{\text{gsw}}+1$ is replaced with $N_{\text{gsw}}$.
We have included loss due to finite success probability at the BSA in the first term, and propagation loss through optical fiber of length $L$ and loss per kilometer $l_f$ in the second term.

For the group switch, we assume a Bene\v{s}'s non-blocking optical switch network \cite{Spanke:87}.
Every photon passes through ($2\lceil \log_2{k}\rceil -1$) elementary $2 \times 2$ switch points within every group switch.
Therefore, the insertion loss of every group switch with radix $k$ is roughly
\begin{equation}
    \mathrm{GroupLoss}(k) \approx \left( 2\lceil \log_2{k}\rceil -1 \right) loss_{\mathrm{2 \times 2}},  \label{formula:grouploss}
\end{equation}
where $loss_{\mathrm{2 \times 2}}$ is the insertion loss of a single $2 \times 2$ switch point.

The actual GroupLoss($k$) and the switching time $t_{\text{gsw}}$ depend on the chosen switch technology.
Our elementary switch points have loss around 0.4 dB, as shown in Tab.~\ref{tab:lossbudget}.
They also suffer from long switching times at the order of miliseconds, with the cost of polarization compensation and ODL being even higher as it has to be done manually in the current experimental setup.
Numerous other switch technologies exist with various trade-offs between insertion loss and switching times (see Appendix~\ref{sec:candidates-appendix}).

Finally, group switch infidelity originates from inaccuracies in polarization stabilization as well as imperfections in BSA measurements. Based on our laboratory measurements, we further assume that effects from the group switch itself are negligible. In this model, the infidelity caused by Q-Fly is independent from the configuration-dependent parameters Tab.~\ref{tab:notation}. Based on our data, the infidelity caused by Q-Fly is about ten percent in our lab, a figure we will reuse here.

\subsection{Computation Time Estimation}

With the above basis, we next evaluate the Q-Fly workload using a performance evaluation tool we have developed.

\subsubsection{Evaluation setup and conditions}

We estimated the total losses depended on actual topologies, with parameter settings shown in Tab.~\ref{table:config_simulation}, along with the actual losses.
We use the insertion loss of a single $2 \times 2$ MEMS switch of 0.4 dB from Tab. \ref{tab:lossbudget}, which means single photon passes through this $2 \times 2$ switch with transmittance of 90\%.

Control is assumed to be centralized with control over all computational resources and network devices; therefore, all operations are executed at the scheduled time and with precisely controlled duration. We assumed durations for local logical one- and two- qubit gates are shorter than remote logical two-qubit gates including communication overhead obtained from Eq. (\ref{formula:loss_Q-Fly_HD}).
Based on the mantra \emph{Every decibel matters}, usually we select the path to minimize the number of hops over group switches. However, sometimes we obtain an earlier operation time by comparing the communication queue length and routing along a longer path.

\begin{table}[]
    \centering
    \caption{Evaluation settings. Number of data qubits per end node is $q$, and the total number of data qubits is $q_{tot} = Nq$. $N_{sw}$ means the number of $2 \times 2$ switches in every group switch as mention about Eq. \ref{formula:grouploss}.  The loss is for the normal shortest path, including all factors.  PL, $\operatorname{PathLoss}_\mathrm{HD}$.}
    \begin{tabular}{lrrrrrrrr}
    \hline
    Exp. & $p$  & $k$  & $g$  & $N$   & $q$  & $q_{tot}$ & $N_{sw}$ & PL [dB] \\
    \hline
    1 & 2  & 4  & 5  & 10  & 13 & 130        & 3         & 7.2               \\
    2 & 6  & 8  & 5  & 30  & 4  & 120        & 5         & 9.9               \\
    3 & 4  & 8  & 9  & 36  & 4  & 144        & 5         & 9.9               \\
    4 & 2  & 8  & 13 & 26  & 5  & 130        & 5         & 9.9               \\
    5 & 14 & 16 & 5  & 70  & 2  & 140        & 7         & 12.7               \\
    6 & 8  & 16 & 17 & 136 & 1  & 136        & 7         & 12.7               \\
    \hline
    \end{tabular}
    \label{table:config_simulation}
\end{table}

\subsubsection{Performance estimation}

\begin{figure}
    \centering
    \includegraphics[width=1\linewidth]{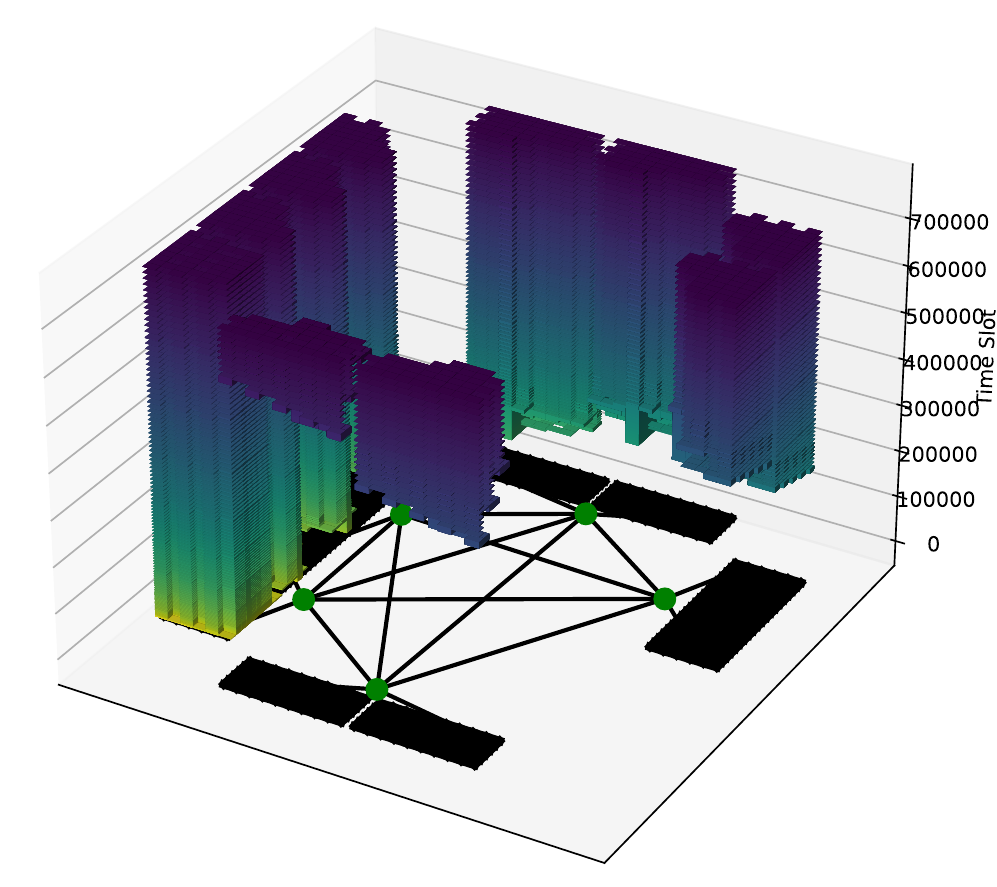}
    \Description[A 3D graph illustrating a network of switches and the stacked states of end-nodes connected to them]{The switches are labeled as S1 to S4 and nodes are represented as planar 3 by 3 square grids. The state of each end-node has stacked along the z-axis, representing the changes over time.}
    \caption{Scheduling of QFT operations on one quasi-half duplex, single-path Q-Fly network configuration. The floor is an abstract representation of the system, with nodes containing logical qubits as the black rectangles, group switches as green dots and group-to-group links as undirected lines.
    Detector cryostats are not drawn. Time is on the vertical axis, where the time slot is defined to be the time to execute a single logical qubit gate.  Each active logical qubit is drawn as a colored block.
    }
    \label{fig:compiled-execution}
\end{figure}

Figure~\ref{fig:compiled-execution} shows spatial and temporal resource occupation for Experiment \#1 in Tab.~\ref{table:config_simulation}, depicted as block objects.
Segments of blocks and lines which have the same color represent single-qubit gates.
Note that due to the large number of gates, the color varies smoothly. 
The yellow colors correspond to the first logical qubit gates executed.
Then, roughly all end nodes act in a clockwise stream, illustrating the essentially sequential nature of the QFT algorithm.

Because CNOT gates using lattice surgery need to occupy resources (tiles) between the control qubit and the target qubit during execution, parallel execution of multiple CNOT gates may be prevented due to competition for resources within each node.
Therefore, the total execution time depends on the network topology and placement of logical qubits within and between nodes.

\begin{figure}
    \centering
    \includegraphics[width=1\linewidth]{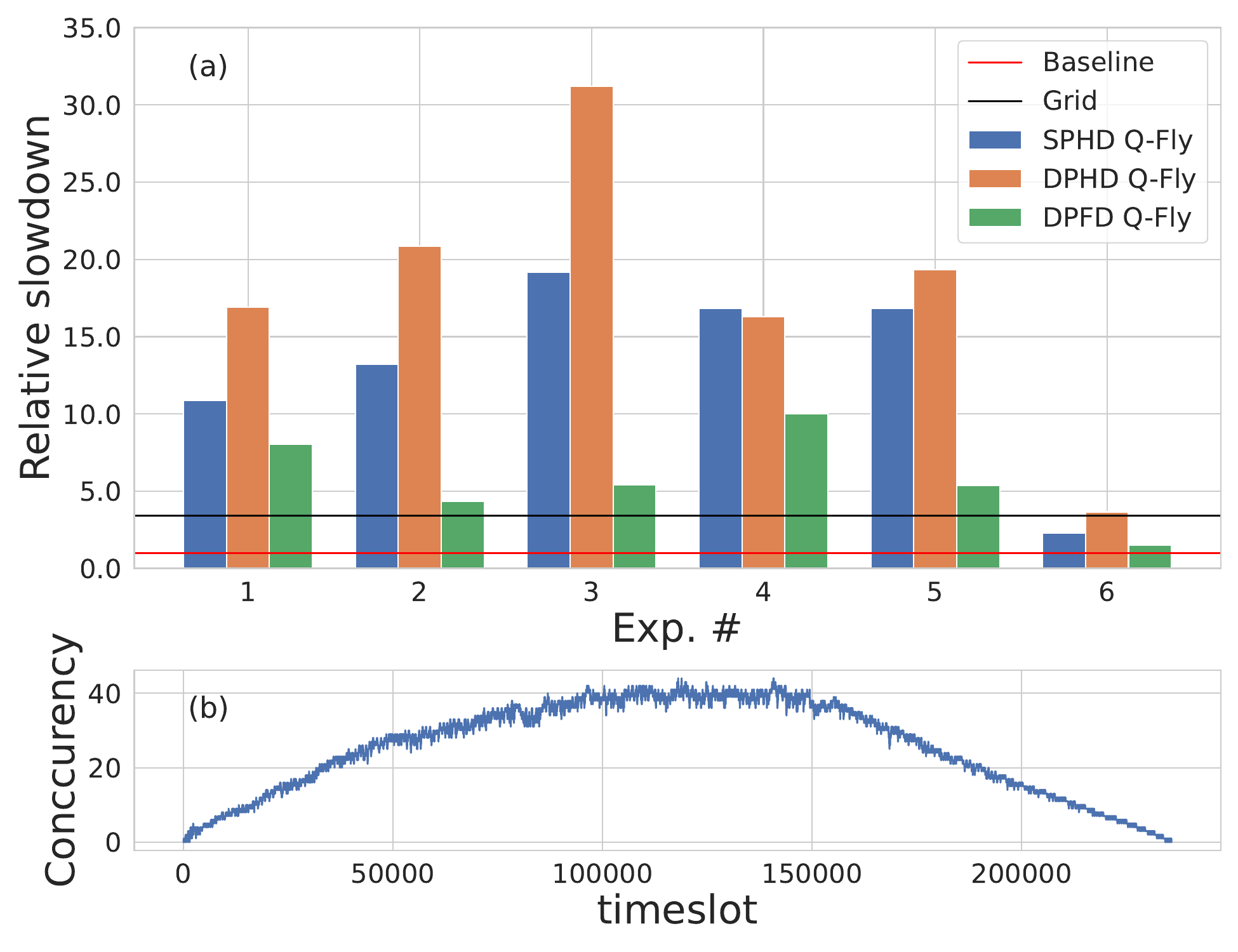}
    \Description[Execution time comparison bar graph.]{Execution time comparison bar graph.}
    \caption{ (a) Execution time comparison. Each bar indicates the execution time for each experiment pattern in Tab. \ref{table:config_simulation}, and color differences are corresponding to Q-Fly's variants in Fig.\ref{fig:q-flies}.
    The red line is the baseline, in which all the logical qubits are in one node, and which is expected to be infeasible in practice since the system is too large.
    Grid shows the performance of a 2-D grid topology where computational nodes are connected to up to four neighboring nodes directly.
    The vertical axis indicates the slowdown factor relative to the baseline.
    The Q-Fly network in pattern 6, where each end node has only one logical data qubit, scores the best and takes only 2.3 times longer to execute than the baseline monolithic case. (b) Concurrency of Exp. \#6 of SPHD Q-Fly showing efficient use of the interconnect leading up to 40 simultaneous active connections between logical qubits during the execution.
     }
    \label{fig:result_evaluation}
\end{figure}

Fig.~\ref{fig:result_evaluation} compares the estimated performance of Q-fly networks shown in Tab.~\ref{table:config_simulation}, a monolithic baseline case, and a 2-D square grid of nodes.
Because the absolute execution time is technology-dependent, we stick to the relative comparison and renormalize all execution times to the monolithic case.  Because the examples here are still of moderate size, the monolithic and grid cases perform well, however, they are impractical physical designs at larger scales.  Moreover, the grid comparison is in some sense unfair to the Q-Fly topologies, since it uses four network interfaces per node compared to one or two for Q-Fly.

The graph shows that the performance depends largely on the network topology.
Experiment \#1 shows use of multi-hop routing and effects of contention for inter-group links.
Due to the 1.4 dB GroupLoss(k=4), the time for one round is 1.3x slower, but with achieved concurrency, overall performance improves.
The best case is Experiment \#6, consisting of 17 groups, each comprising a switch with radix of 16 and 8 end nodes, where each end node holds only one logical data qubit.
This setting also gives high concurrency, with up to 40 simultaneously active connections, but requires the largest number of fibers and switch ports among our SPHD configurations.
This pattern is the fastest because the competition for resources to execute multiple lattice surgeries in parallel is avoided thanks to the non-blocking switch design, and all CNOT gates can be executed in parallel unless the same logical qubits are used in the CNOT gates simultaneously.

The DPHD variant provide no sustainable advantage over the simpler, cheaper SPHD, due to the higher group switch loss.  DPFD, by avoiding some switches, provides the highest performance, not far off of monolithic designs, but at significant additional hardware cost.

\section{Discussion}
\label{sec:discuss}


With our proposed Q-Fly topology now evaluated in light of our prototype network for one of the most important subroutines, we turn our attention to the next steps for deploying Q-Fly in production systems.

The end-to-end performance of our prototype implementation is only a few \SI{}{Hz} on average. The losses in Tab.~\ref{tab:lossbudget} total to \SI{> 60}{dB}  (depending on which nodes are to be entangled) and the BSA success probability is $1/4$.  This gives a success rate of less than one in a million against a trial rate of \SI{1}{GHz}.
 Another main reason for the low rate is that the SPDC entanglement generation probability is set to $\sim \SI{10}{\%}$ for each attempt. Over two hops, that low probability is squared.  Employing technology such as cavity enhancement of photon emission from atoms could possibly lead to success probability improvements of an order of magnitude or more; however, this approach lengthens photon emission time and correspondingly reduces the trial rate, so the net effect must be engineered carefully~\cite{krutyanskiy2023entanglement}. Another factor decreasing the performance of the network is the switching time to change the state of the group switch. Integrated photonic circuits will be a good option when the state changes of the group switches occur at a high rate.  

Another option for building a high-performance link is to use one-photon interference, which suffers photon loss probability $\sqrt{\eta}$ along a path where two-photon interference suffers in $\eta$~\cite{Lucamarini2018overcoming}. 
About \SI{40}{dB} improvement would be expected in our case, 
at the cost of far more severe phase stabilization demands on the interferometer; path lengths must be stabilized to much less than the wavelength of the photons, in sharp contrast to the looser requirements for two-photon interference in our current system. 

Sharp-eyed readers may have noted that Fig.~\ref{fig:q-fly-experiment} uses entangled photon pair sources as part of each end node, but Fig.~\ref{fig:q-flies} focuses on computational end nodes that can emit photons entangled with their memories.  Here, we are using an EPPS as a stand-in for a computational node as we focus on the network.  The best way to use EPPS sources deserves further study, especially with respect to methods of connecting nodes with memory to the network~\cite{soon2024implementationanalysispracticalquantum}.

Integration of atomic memories such as trapped ions or silicon vacancies in diamond into our proposed interconnect architecture can be done in a straightforward fashion. These memories emit light in the 750-900nm range \cite{krutyanskiy2023telecom,knaut2024entanglement}, meaning transduction may not be necessary for data-center scale interconnects, where losses are dominated by coupling and switching rather than propagation loss in the fiber even at these wavelengths. However, interfacing with superconducting qubits will require transduction from microwave to telecom wavelengths~\cite{Han2021microwave}. Our proposed architecture is fully compatible with transduction, and our analysis can be extended in a straightforward fashion by incorporating the transduction success probability and impact on fidelity.

Logical gate rates in the 1-10kHz range will enable a series of applications.  For a 130-qubit QFT, sufficient for many problems, as in Fig.~\ref{fig:result_evaluation}, a 1kHz logical gate time would give an execution time of about four minutes.  Achieving this rate will require about four orders of magnitude improvement in performance, which matches the 40dB link performance improvements proposed above.


Here, we have proposed three variants of the Q-Fly architecture. Further analysis of all three variants with quantum computing workloads and their traffic patterns (e.g., arithmetic, Hamiltonian simulation, graph state creation, oracles for other problems, or the decoder for DQI~\cite{jordan2024dqi}) is necessary to determine optimal designs for datacenter-scale quantum networks. The relative number and destination of group switches, arrangement of BSAs, and composition of end nodes can all be adjusted, and all affect both the performance and construction cost of such networks.


\section{Conclusion}

We proposed Q-Fly, a quantum interconnect for high-performance connectivity for distributed quantum computation, and evaluated our implementation of many of the elements of a full interconnect stack, from the physical level through compiling distributed programs.
We developed and demonstrated an elementary building block of the Q-Fly network, composed of three optical end nodes and an optical quantum switch node, integrating quantum and classical hardware and software.
We evaluated the performance of the Q-Fly network with a distributed quantum computing performance evaluation tool developed for this work with a QFT circuit.
We discussed the applicability of our demonstrated network to actual distributed quantum computation with end nodes with stationary qubits.

We have found that quantum systems must be designed cross-layer; optimization of subsystems in isolation is not sufficient.  Loss is critical, but can be managed by adopting a low-diameter Dragonfly variant and balancing group size against application traffic locality.  Care must be taken as increasing group size increases intra-group loss.
Polarization and synchronization challenges argue for integrated group switch design, which may encourage nanophotonic chip development.
The physical integration of multi-detector cryostats is well suited to the Dragonfly group structure as adapted into Q-Fly and will scale comfortably in packaging and floor planning.
Our resource estimation shows that, with the proper interconnect structure providing redundant paths, nodes configured with a single available application logical qubit can outperform seemingly more powerful nodes, due to contention for access to the node's communication qubit.  The interconnect is a critical resource, not an afterthought.
Industrial roadmaps have suggested that single-system limits may be reached before the year 2030.  To reach the performance needed for fault-tolerant quantum multicomputers, interconnect development has a challenging but achievable roadmap ahead.


The goal of this work is far more ambitious than just demonstration of a one-off prototype.  We expect portions of our technology to be used in development of much larger, production systems, and to serve as a basis for continued evolution. Even more importantly, we intend for the protocols and network architecture to be standardized~\cite{rfc9340}, much as the interconnects for classical supercomputers are, and are working toward publication of specifications that will advance interoperability and the development of commercial products.

\begin{acks}

This work was supported by JST [Moonshot R\&D Program] Grant Number [JPMJMS226C] (all authors), and Grant Numbers [JPMJMS2066] (RI) and [JPMJMS2061] (RDV).  YU thanks the RIKEN Special Postdoctoral Researcher Program. The authors thank Takao Tomono, Masahiro Takeoka, Tomoyuki Horikiri, Teruo Tanimoto, and Yuya Kawakami for technical advice and student support; Kent Oonishi, Kaori Nogata, Kaori Sugihara, Saori Sato and junsec for logistical support; and the many members of QITF, the Nagayama Moonshot, AQUA, and WIDE for technical collaborations and support.

\end{acks}

\bibliographystyle{ACM-Reference-Format}
\bibliography{references,references-local}

\appendix

\section{Tuning the Prototype}
\label{sec:tuning-appendix}

This appendix discusses some of the processes involved in tuning the prototype to produce the data presented in the main body of the paper, as well as in the next appendix.

\begin{figure}[t]
    \centering
    \includegraphics[width=0.8\columnwidth]{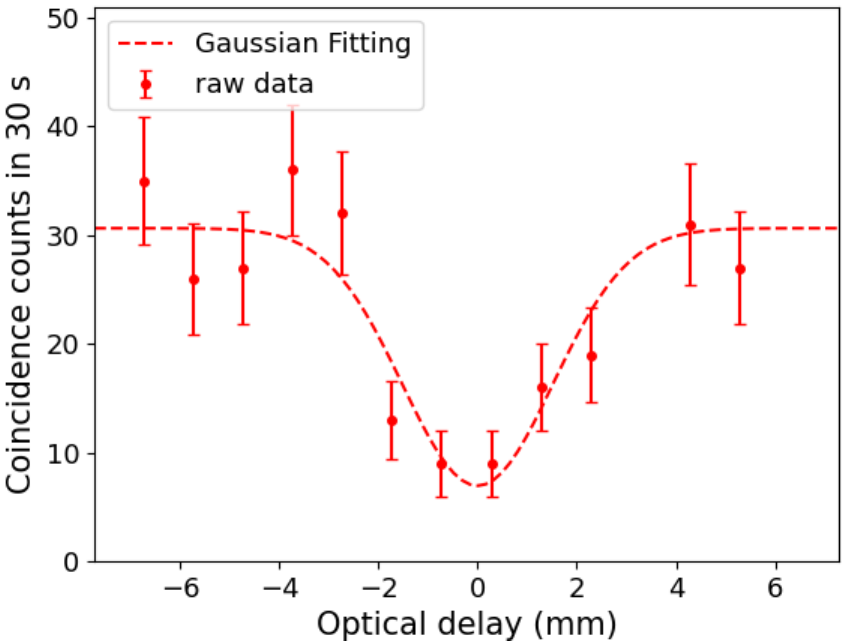}
    \Description[This is the result of Hong-Ou-Mandel experiment shown in a diagram in red color.]{The vertical axis represents the number of coincidence counts measured over 30 seconds ranging from 0 to 45.  The horizontal axis is labeled as optical delay measured in millimeter and ranges from  minus 4 millimeter to plus 4 millimeter. The diagram shows 12 spots illustrating the actual coincidence counts in different optical delay lengths. A dashed curved, based on Gaussian fit of the raw data points, is also included.}
    
    \caption{HOM dip between photons $A_1$ and $B_1$. 
    This demonstrates that we can properly synchronize their arrival times at the BSA 
    with the dip where the non-classical HOM visibility larger than 0.5 indicates the two photons are highly indistinguishable. The FWHM is \SI{3.56}{mm}, 
    which shows that wavelength-order precision of the control is not required for the ODL. 
    }
    \label{fig:HOM dip}
\end{figure}

We performed the Hong-Ou-Mandel~(HOM) experiment to adjust the optical delays of the photons arriving from the different nodes~\cite{hong-PhysRevLett.59.2044}. 
For example, for the HOM interference between the photons at modes $A_1$ and $B_1$, 
the H-polarized photons are mixed by measuring the V-polarized photons at modes $A_2$ and $B_2$. 
The experimental result of the HOM dip is shown in Fig.~\ref{fig:HOM dip}.
From the best fit to the experimental data with Gaussian functions, 
the visibility of the HOM interference between the photons from modes $A_1$ and $B_1$ is estimated to be $0.77\pm 0.08$.
Visibility is given by the difference between maximum and minimum recorded coincidence counts and renormalized by their sum.
Similar experiments related to modes $(B, C)$ and $(C, A)$ were performed. 
The visibilities of the HOM experiments are $0.68\pm 0.04$ and $0.67\pm 0.18$, respectively. 

\section{Details of Scalability Experiments}
\label{sec:app-scalability}
To demonstrate the scalability of Q-Fly, we designed an additional experiment capable of adding more switches to the optical path under software control, emulating the loss of a higher-radix, deeper group switch. To achieve this, several available switch components were stacked such that the output of one switch component was connected to the input of the next. This looping configuration enables us to introduce $2L$ new switching layers by adding $L$ switch components.

\begin{figure}[hbt]
    \centering
    \includegraphics[width=0.45\textwidth]{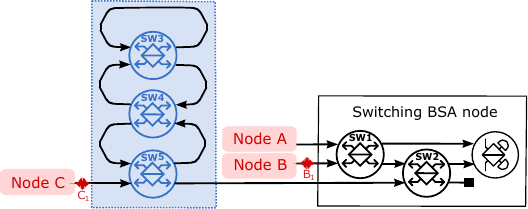}
    \Description[]{}
    \caption{Extending our prototype network by adding looped switch components allows us to analyze the impact of higher-radix, higher-loss group switches on the performance of the system.  Under software control, the photon from node $C$ can be made to pass through 2, 4, 6, or 7 switch components.}
    \label{fig:LoopingConfiguration}
\end{figure}

The general setup of the extended experiment is shown in Fig.~\ref{fig:LoopingConfiguration}. The detailed hardware configuration of end-node and switching BSA is the same as the baseline experiment (Fig.~\ref{fig:q-fly-experiment}). The only difference is that an additional optical delay line is concatenated with the first ODL to extend our ability to adjust the path length from 200 mm to 400 mm, covering all possible combinations of arrival times. 

The new emulated group switch, consisting of three looped switch devices, is placed between one of the end nodes (node $C$ in this experiment) and the switching BSA node. Depending on the status of each switch device, the photon path $C_1$ traverses a different number of switch layers before reaching the switching BSA node, which is connected to the last switch in the chain. Four meaningful configurations are possible for the three switches (SW$_5$, SW$_4$, SW$_3$): $(1, x, x)$, $(2, 1, x)$, $(2, 2, 1)$, and $(2, 2, 2)$, where states 1, 2, and $x$ represent bar, cross, and don't care, respectively. Taking into account these configurations, along with the state of the switch device in the switching BSA node (SW$_2$), the photon pair in path $C_1$ passes through 2, 4, 6, or 7 switch devices.

\begin{figure}[htb]
    \centering
    \includegraphics[width=0.5
    \textwidth]{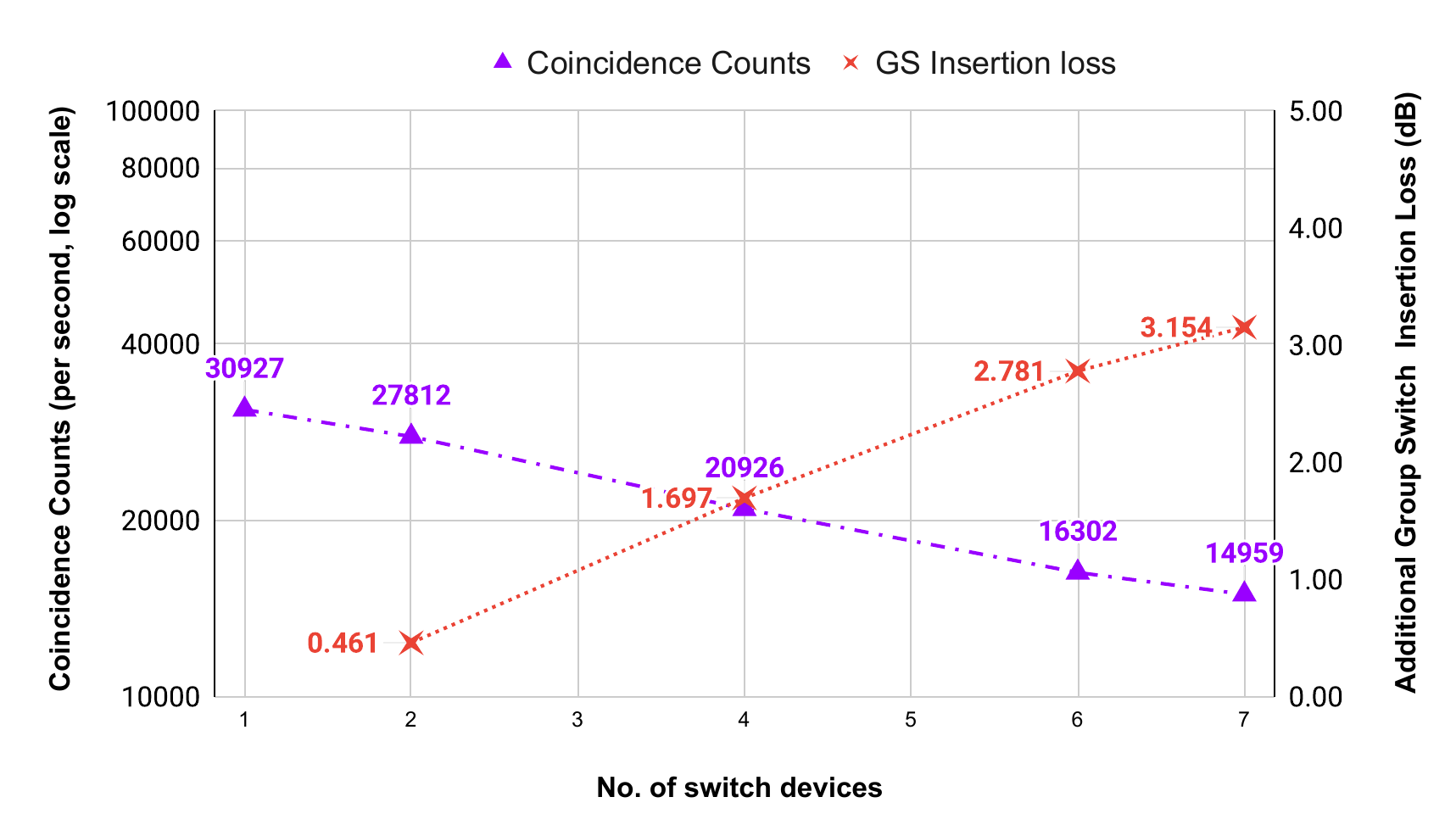}
    \Description[]{}
    \caption{Adding switching layers by changing the switch settings in Fig.~\ref{fig:LoopingConfiguration} reduces the single-photon coincidence counts generated by node $C$ (left axis), from which we calculate the associated insertion loss (right axis).  This data was taken with an H polarization filter on one arm, and no filter on the other, and does not correspond exactly to Bell pair generation rate.}
    \label{fig:GroupSwitch-IL-Analsis}
\end{figure}

We first analyze the performance of the entanglement generated in node $C$ by measuring the coincidence counts of its photon pairs as photon $C_{1}$ passes through different switch layers. For this setup, the emulated switch group and SW$_1$ of the Switching BSA node are used. In the first step, we bypass the emulated group switch and send photon $C_{1}$ directly through SW$_1$ of the Switching BSA. Next, we insert the emulated switch group and apply four different configurations. We then measure the coincidence counts of the photon pair $(C_{1}, C_{2})$. The coincidence count rate (per second) and the insertion loss relative to the bypassed configuration are shown in Fig.~\ref{fig:GroupSwitch-IL-Analsis}. 

\begin{figure}[htbp]
    \centering
    \includegraphics[width=0.45\textwidth]{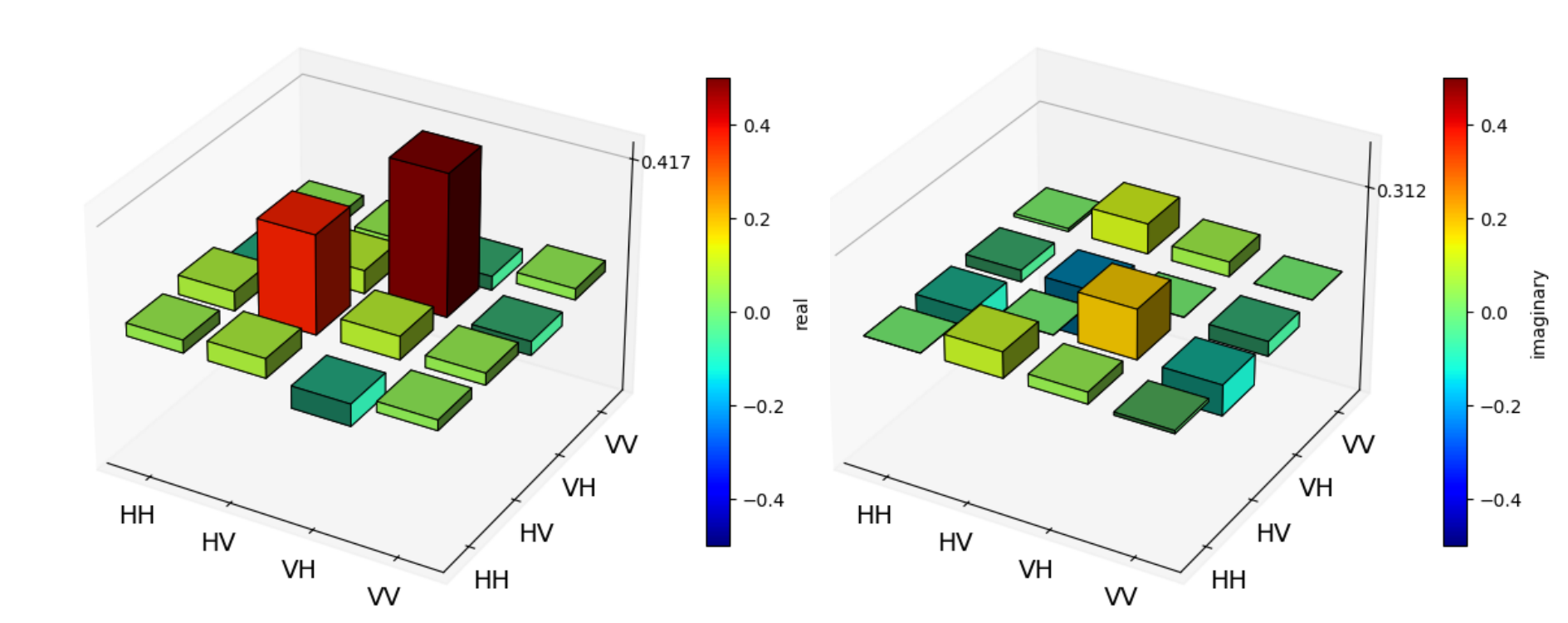}
    \Description[]{}
    \caption{Real (left) and imaginary (right) parts of the reconstructed density matrix  $\rho_{B_2C_2}$ for the longest-path (highest switch-depth) configuration after entanglement swapping}
    \label{fig:BC222_densitymatrix}
\end{figure}

To evaluate the quality of entanglement swapping between nodes in the setup, we select one of the other nodes as a partner for node $C$. In this experiment, we chose node $B$ to be the entanglement partner of node $C$. For each switch configuration, we first perform a Hong-Ou-Mandel experiment, followed by quantum state tomography. As one of the results, the reconstructed density matrix corresponding to the longest switch depth with the looped switch state $(2,2,2)$ is shown in Fig.~\ref{fig:BC222_densitymatrix}.

The results of all configurations are shown in Table~\ref{tab:ScalabilityDemonstration}. The entanglement rate is based on this formula~\cite{tsujimoto2018high}: 

\begin{equation}
    \textrm{Rate}_{\textrm{Ent}} = \sum_{\textrm{Basis}\in\{HH,HV,VV,VH\}} \frac{\textrm{FourFold}_{\textrm{Basis}}}{\textrm{MeasurementTime}_{\textrm{Basis}}}.
\end{equation}

\begin{table*}[h]
    \centering
    \begin{tabular}{| c || c | c | c | c | c | c |}
         \hline
         State of Switches  & No. of Switches  & Visibility & Fidelity & Entanglement  & Purity  \\ 
         (SW$_5$,SW$_4$,SW$_3$)  & $B_1$ and $C_    1$ pass  &   &  & Rate (bpps) &  \\   
         \hline
         (1,x,x) & 3  & 0.66 & 0.60 $\pm$ 0.07 &  0.217 & 0.53 $\pm$ 0.05\\ \hline
         
         (2,1,x) & 5  & 0.76 & 0.67 $\pm$ 0.09 &  0.153 & 0.53 $\pm$ 0.09\\ \hline
         
         (2,2,1) & 7  & 0.81 & 0.76 $\pm$ 0.1 &  0.147 & 0.62 $\pm$ 0.12\\ \hline

        (2,2,2) & 8  & 0.73 & 0.72 $\pm$ 0.07 &  0.147 & 0.61 $\pm$ 0.06\\ \hline
    \end{tabular}
    \caption{The experimental results of HOM and Entanglement Swapping between node B and node C in the extended prototype with the emulated group switch}
    \label{tab:ScalabilityDemonstration}
\end{table*}
\section{Group Switch Candidate Technology Analysis}
\label{sec:candidates-appendix}

\begin{figure*}[ht]
    \centering
    \includegraphics[width=0.95\textwidth]{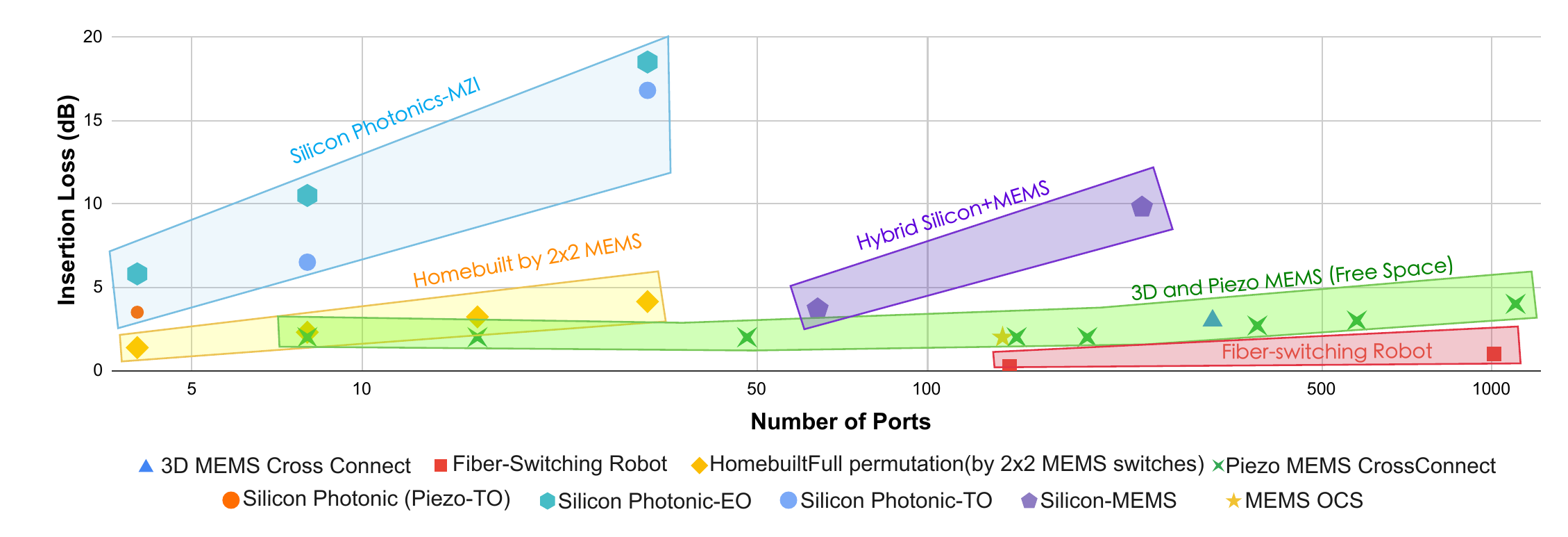} 
    \Description[The diagram illustrates insertion loss in various candidate switch technologies]{The diagram's horizontal axis represents the number of ports on a logarithmic scale, ranging from zero to 1000. The vertical axis is insertion loss in the scale of decibel ranging from zero to 20. The figure categorized different switch technologies or fabricated chips into five different categories: Silicon photonic category, using Mach-Zehnder Interferometer components, is positioned on the left-top side in blue color. The second category, labeled "Homebuilt by 2x2" and shown in yellow, is located in the left side. The "Hybrid Silicon+MEMS" category, shown in purple, is positioned in the middle of the diagram. The "3D and piezo MEMS" category shown in green, spans from the left-bottom to the right-bottom side of the diagram. The fifth category, "Fiber switching robot", shown in red color, is located at the right-bottom side of the diagram}
    \caption{Candidate group switches: insertion loss in all-optical switches using different designs and technologies. Each point corresponds to a specific design or product enabling all-to-all full permutation and non-blocking mapping of inputs to outputs. These range from silicon photonic chips with Mach-Zehnder interferometer (MZI) components utilizing electro-optic (EO)~\cite{yang2010non,dupuis20208,qiao201732}, thermo-optic (TO)~\cite{suzuki2014ultra, tanizawa2015ultra} or piezo-TO~\cite{dong2022high} actuation mechanism to MEMS (3D cross connect~\cite{kaman2007compact, Calient}, piezo cross-connect~\cite{Polatis}, and MEMS OCS with camera based closed-loop alignment~\cite{liu2023lightwave}). Other designs include hybrid silicon-MEMS~\cite{seok2016large, seok2019240} and fiber-switching robots~\cite{kewitsch2009large,Telescent}. In the homebuilt category, the loss is assessed based on our measurements of loss of basic 2$\times$2 MEMS switches and a standard Bene\v{s} interconnect topology.}
    \label{fig:switch-IL-comparision}
\end{figure*}

Here, we detail a number of switch technologies and compare them with our groups constructed group switch.
As mentioned in Sec.~\ref{sec:2-node}, our switch points have loss around 0.4 dB and switching times of few miliseconds.
For example, fabrication of Mach-Zehnder switching components on silicon photonic chips is a promising technology, offering several benefits including a switching time ranging from nanoseconds in electro-optical switches~\cite{dupuis20208,qiao201732} to microseconds in thermo-optical switches~\cite{suzuki2014ultra, tanizawa2015ultra}. However, as shown in Fig.~\ref{fig:switch-IL-comparision}, such switches suffer from significant fiber-to-fiber insertion loss that will grow as the group switch size increases.

In contrast, cross-connect MEMS technology enables high-radix switches with low fiber-to-fiber insertion loss~\cite{Calient,Polatis,kaman2007compact,liu2023lightwave}, but with much slower switching time, typically on the order of a hundred milliseconds. These differing features have motivated design and fabrication of hybrid Silicon-MEMS devices, such as MEMS actuated vertical adiabatic couplers ~\cite{seok2016large, seok2019240}, to take advantage of the best qualities of each technology. This can increase switch scale while reducing insertion loss and switching time.

Finally, fiber-switching robots~\cite{kewitsch2009large,Telescent}, designed for automation of topology management, offer lower insertion loss. However, their slow reconfiguration time, on the scale of minutes, limits their effectiveness in dynamic networks. Further performance improvements would be required to enable their use in quantum interconnection networks.

As illustrated in Fig.~\ref{fig:switch-IL-comparision}, commercially available switches today have a capacity of up to approximately $k=576$, with experimental switches up to $k=1100$~\cite{kim03:_1100_port_mems}.

\end{document}